\documentclass[sigconf,nonacm]{acmart}

\setcopyright{none}
\makeatletter
\@for\@tmpn:=1,2,3,4,5,6,7,8\do{\expandafter\let\csname typeset@author\@tmpn\endcsname\@firstofone}
\makeatother
\renewcommand{\footnotetextcopyrightpermission}[1]{}
\renewcommand{\footnotetextauthorsaddresses}[1]{}

\usepackage[utf8]{inputenc} 
\usepackage[T1]{fontenc}    
\usepackage{hyperref}       
\usepackage{url}            
\usepackage{amsfonts}       
\usepackage{nicefrac}       
\usepackage{microtype}      
\usepackage{xcolor}         
\usepackage{colortbl}       
\usepackage{siunitx}
\usepackage{graphicx,subcaption}
\usepackage{amsmath}
\usepackage{enumitem}
\usepackage{multirow}
\usepackage{balance}
\usepackage{listings}
\usepackage[ruled,vlined]{algorithm2e}
\usepackage{amsmath}

{\definecolor{codegreen}{rgb}{0,0.6,0}
\definecolor{codegray}{rgb}{0.5,0.5,0.5}
\definecolor{codepurple}{rgb}{0.58,0,0.82}
\definecolor{backcolour}{rgb}{0.95,0.95,0.92}
\lstdefinestyle{mystyle}{
    backgroundcolor=\color{backcolour},   
    commentstyle=\color{codegreen},
    keywordstyle=\color{magenta},
    numberstyle=\tiny\color{codegray},
    stringstyle=\color{codepurple},
    basicstyle=\ttfamily\footnotesize,
    breakatwhitespace=false,         
    breaklines=true,                 
    captionpos=b,                    
    keepspaces=true,                 
    numbers=none,                    
    numbersep=5pt,                  
    showspaces=false,                
    showstringspaces=false,
    showtabs=false,                  
    tabsize=2
}

\author{Seunghyun Oh$^{1*}$, Hirotaka Hiraki$^{2*}$, Shuyue Stella Li$^{1}$,
Yulia Tsvetkov$^{1}$,\\
Shyamnath Gollakota$^{1}$\\}
\makeatletter
\g@addto@macro\@authornotes{\stepcounter{footnote}\footnotetext{Both authors contributed equally to this research.}}
\makeatother
\affiliation{
\institution{$^1$Paul G. Allen School of Computer Science and Engineering, University of  Washington, WA, USA\\}
\institution{$^2$Graduate School of Interdisciplinary Information Studies, The University of Tokyo, Bunkyo, Tokyo, Japan}
\country{}
}

\lstset{style=mystyle}}

\definecolor{JarvisPaletteRed}{HTML}{D84E34}
\definecolor{JarvisPaletteYellow}{HTML}{E7BF59}
\definecolor{JarvisPaletteGray}{HTML}{929292}
\definecolor{JarvisPaletteGreen}{HTML}{95D365}
\definecolor{JarvisPaletteBlue}{HTML}{5F9DF1}

\colorlet{JarvisGapBlue}{JarvisPaletteBlue}
\colorlet{JarvisWrongRed}{JarvisPaletteRed}
\colorlet{JarvisSelfCorrectionYellow}{JarvisPaletteYellow}
\colorlet{JarvisResolvedquestionGreen}{JarvisPaletteGreen}
\colorlet{JarvisUnanswerableGray}{JarvisPaletteGray}

\newcommand{\xref}[1]{\S\ref{#1}}

\newcommand{\squishlist}{\begin{itemize}[itemsep=1pt,parsep=2pt,topsep=3pt,partopsep=0pt,leftmargin=0em, itemindent=1em,labelwidth=1em,labelsep=0.5em]}
\newcommand{\squishend}{\end{itemize}}

\setlist[itemize]{leftmargin=*}

\begin{document}

\title{{Jarvis: A Proactive Speech Agent for Multi-Party Conversations}}

\begin{teaserfigure}
  \includegraphics[width=\textwidth]{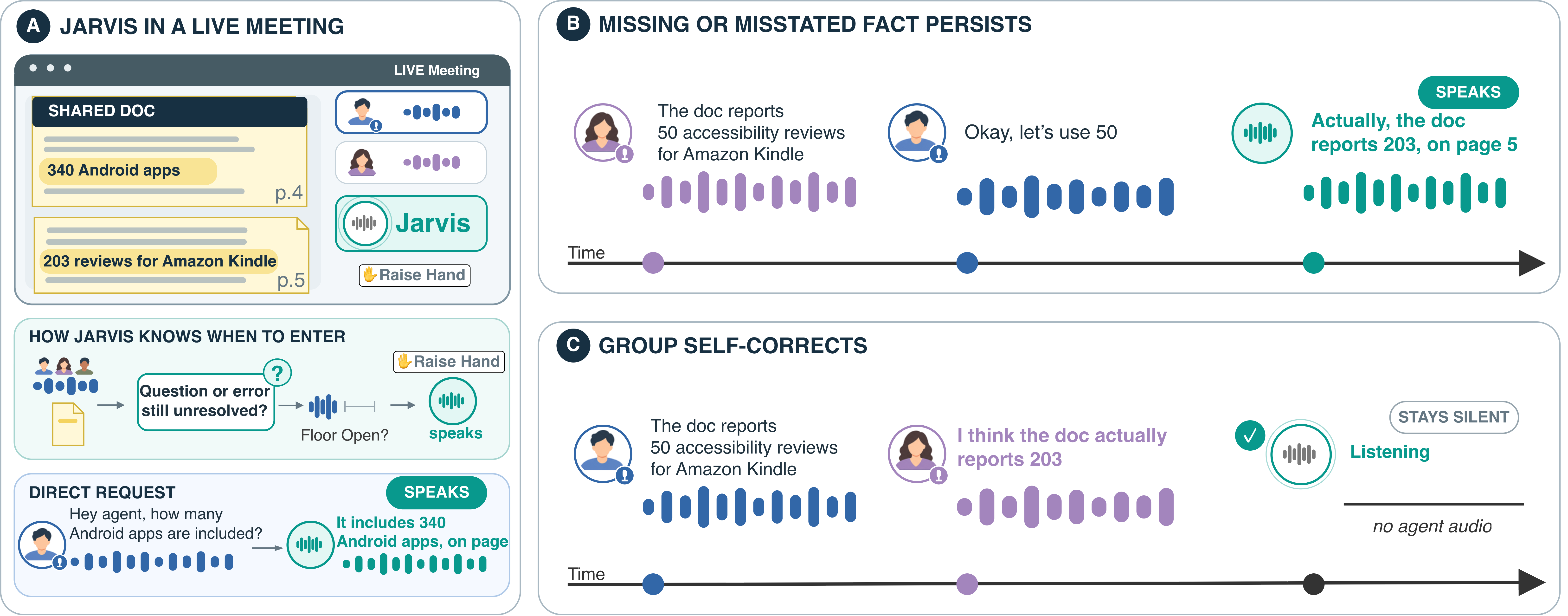}
  \vskip -0.15in
  \caption{\textbf{Jarvis in a live multi-party meeting.}
(A) Jarvis joins the meeting as a participant, follows the conversation against a shared document, and can answer direct questions. For proactive intervention, it has two modes: 1) raises its hand to seek permission or  2)  waits until no participant holds the floor and speaks directly. (B) If a document-grounded missing or misstated fact remains unresolved, Jarvis speaks to correct it and cites the relevant page. (C) If the group corrects the error first, Jarvis   remains silent.}
  \Description{Overview of Jarvis in a live Zoom meeting. Jarvis follows a shared document and can answer direct requests. In one example, participants continue discussing an incorrect number, so Jarvis raises its hand, waits for an open conversational floor, and provides the correct number with a page citation. In another example, a participant corrects the number before Jarvis  remains silent.}
  \label{fig:teaser}
\end{teaserfigure}

\renewcommand{\shortauthors}{Seunghyun Oh, Hirotaka Hiraki, et al.}


\begin{abstract}
Speech agents are reactive and dyadic: they speak when spoken to, and to one person at a time. We ask what it takes for a speech agent to instead take part in a conversation among several people and speak up only when it can help. We introduce Jarvis, a real-time proactive speech agent that audibly participates in multi-party human conversations. Grounded in a document shared beforehand, Jarvis follows the discussion and intervenes when the group misses or misstates a fact and does not correct itself within a few turns. We make three contributions: a problem setting based on epistemic breakdowns that makes proactive intervention measurable, realized as \texttt{CHI-180-proactive}, a synthetic multi-party dataset seeded with known gaps, errors, and self-corrections; a proactive backbone that harnesses a small, open-weight model with deterministic checks and grounds every claim in a source sentence; and interaction techniques for taking the floor in live speech and showing the cited evidence on screen. On \texttt{CHI-180-proactive}, Jarvis is correct on most events it addresses and stays silent 97\% of the time when the group resolves an issue itself. A live study with 23 participants confirms these trends with real-time interventions.
\end{abstract}

\maketitle

\section{Introduction}

Imagine a meeting where everyone is working with a 15-page contract, but nobody remembers every detail. Someone asks what the contract says about data retention. Nobody knows, so the question gets pushed to the next meeting. Later, someone misstates the sample size, and the discussion starts building on the error. Now imagine an agent that has read the contract, answers questions on the spot, and listens to the conversation. When someone misses or misstates a fact, it first gives the other participants a chance to catch and correct it. If they do not, it speaks up. Such a {\it proactive} speech agent could be useful in lab meetings, diligence calls, design reviews.

Today's speech agents cannot play this role for three key reasons. First, they are primarily reactive: from smart speakers to LLM voice modes, they speak only when addressed, e.g., ``Hey Alexa.'' But some of the most useful moments for help are ones no one thinks to ask for, such as when a fact has been misstated without anyone realizing it. Second, they are designed for dyadic interaction~\cite{defossez2024moshi,syncllm}: their turn-taking, endpointing, and barge-in mechanisms assume a single user is addressing the agent, so an utterance is generally treated as a turn that warrants a response. A human-conversation agent must work differently. It is listening, but without being explicitly addressed, it  must decide when the conversation needs intervention, give participants a chance to correct themselves, and then enter an ongoing conversation without derailing it. Third, it must remain in the background and keep humans at the center, speaking only when necessary rather than competing for attention. Neither reactive assistants, dyadic dialogue systems, nor in-ear whispering assistants~\cite{chen2025llamapie,10.1145/3613904.3642450} provide the interaction model or evaluation framework needed for this setting.

We introduce Jarvis, to our knowledge, the first real-time  proactive speech agent that audibly participates in multi-party human conversation. Designing and evaluating this requires addressing three core research questions:

\begin{itemize}

\item {\it{{\bf RQ1:} Should the proactive agent speak, and what should it say?}} At each turn, the agent must decide whether to say anything and, if so, what. Unlike a reactive assistant where humans ask for explicit assistance, the multi-party conversation can continue without any intervention from the proactive speech agent.   So the agent has to identify the rare moments when its intervention actually helps and is not considered a distraction  by the human participants.

\item {\it{{\bf RQ2:} When and how should it take/hold the floor?}} Unlike text-based proactive agents, where a chat message can be skimmed, deferred, or ignored, a spoken intervention consumes shared airtime and displaces whoever might speak next. The agent can neither cut across someone mid-sentence nor sit on an important point until the moment has passed. And if a participant does interrupt it, the agent must decide whether to keep the floor or yield, and what to do with the intervention it was partway through delivering.

\item {\it{{\bf RQ3:} How do we measure whether proactive intervention worked?} } Proactive behavior has been difficult to evaluate because open-ended conversation has no answer key. Existing human-conversation datasets do not fill this gap: AI speech agents should be secondary participants in human conversations, taking a back seat to the humans, yet no datasets capture this setting. Without a well-defined problem setting, whether an interjection was correct, well-timed, or needed at all becomes a matter of opinion. We therefore need problem settings and datasets in which both correctness and timing are objectively decidable, allowing us to evaluate what an agent says and when it says it against something firmer than opinion.

\end{itemize}

Our paper addresses these questions and makes the following conceptual and technical contributions.
\begin{itemize}

\item {\it Multi-party setting and dataset that make proactive intervention measurable.}
Proactive behavior is hard to evaluate because open conversation has no answer key. To make proactive intervention decidable by construction, we target \emph{epistemic} breakdowns rather than social ones. Specifically, Jarvis identifies moments where the multi-party group is missing or misstating something checkable. To do this, we  ground each conversation in a shared document (e.g., financial report or academic paper) that is shared apriori with Jarvis before the start of the conversation. We also  introduce \texttt{CHI-180-proactive}, a dataset created from 180 CHI papers, each paired with a synthesized multi-party discussion into which we plant 24 known events like knowledge gaps, errors, self-corrections, and distractors. In total the dataset has  4{,}320 events over 50{,}400 turns, each annotated with a fixed correct action. This turns proactive intervention into a detection task with measurable accuracy and timing, and we will release it as a research benchmark for proactive conversational agents.

\item {\it A deterministic harness of small models that decides whether to speak and what to say.}
We harness a  small, open-weight language model through several stages with deterministic checks, so the whole system can be self-hosted and no single model output is trusted on its own. At each turn one model classifies whether the discussion is missing or misstating something the document can settle and  another checks whether the participants have already corrected it. A  deterministic gate reconfirms and speaks only once a problem has persisted, so the agent defers to human self-repair before intervening. Every spoken claim is grounded in a document sentence rather than the model's own knowledge (\S\ref{sec:proactive}). Our evaluation shows that in this document-grounded setting our harness of small models is enough and is fast enough to run in real time.

\item{\it Interaction techniques for taking and holding the floor.} Our goal is for 
Jarvis to join a live conversation like a considerate participant. To achieve this, we explore  three interaction methods. First, rather than speaking as soon as it has something to say, it queues the intervention and waits for an opening; if someone begins talking while it is speaking, it stops at once and decides whether to replay, resume, or drop what it was saying. Second, for groups that want tighter control, it instead raises its hand and speaks only once someone verbally gives it the floor. Third, alongside the spoken answer it shows visual evidence, placing the cited source on screen so participants can verify a claim rather than take it on the agent's word (see~\xref{sec:interact}).

\end{itemize}

Rather than build a Wizard-of-Oz prototype, we address the underlying systems and latency challenges, deploy a proof-of-concept Jarvis on Zoom as an additional participant, and run it live in multi-party conversations. 

Our key findings are as follows:
\begin{enumerate}

\item Our   \textsc{Qwen3.6-35B-A3B} harness, with 3B active parameters, is competent in our document-grounded setting: on  \texttt{CHI-180-proactive}, Jarvis provided correct or partially correct information with  95-97\% accuracy.

\item Jarvis defers to the group and intervenes when needed. After  participants self-correct ($N{=}720$) or answer their own question ($N{=}720$), it stays silent in 97\% of cases in both. When there is an information gap ($N{=}1800$) or a document-specific  incorrect statement ($N{=}720$), it intervenes in 91\% and 77\% of cases; the latter is lower because many incorrect statements import a fact from another document, which Jarvis cannot verify against its own source and hence stays silent.  When it does intervene, it acts within 2-3 turns of the event in 98\% of cases.

\item In live Zoom sessions with 23 participants, where Jarvis must also cope with transcription errors from the speech pipeline, these results carry over to real people: 85\% of its interventions arrived within two turns of the problem, and 163 of 164 carried correct or partially correct information. Together, Jarvis and the participants resolved 74\% of planted errors, 72\% of spontaneous errors, and 83\% of information gaps.

\item When asked to choose, participants preferred an agent that raises its hand before speaking to one that speaks directly, and overwhelmingly preferred having the cited source shown on screen (21 of 23), which let them check claims and trust the answers.
\end{enumerate}

Taking a step back, we show that real-time proactive participation in spoken conversation can be   designed  and carefully measured against ground truth, establishing a research foundation for future proactive speech agents.

\section{Related Work}

 
\vskip 0.05in\noindent{\bf Proactive Agents and the When-to-Speak Problem.} 
Mixed-initiative systems~\cite{horvitz1999principles} and just-in-time agents such as the Remembrance Agent surface relevant information from observed activity~\cite{rhodes1996remembrance}, but deliver it on screen rather than into the conversation. Interruption research showed that poorly timed interventions carry cognitive costs~\cite{mark2008cost,iqbal2008effects} and developed models for predicting interruptibility and deferring notifications~\cite{horvitz2003learning,horvitz2004busybody,horvitz2005balancing}; these focus on individual notifications, not spoken contributions. More recent work on proactive voice assistants finds a similar ``proactivity dilemma'' in balancing helpfulness and intrusiveness, particularly in multi-party settings~\cite{zargham2022proactivity,oh2024better}, but remains largely formative and Wizard-of-Oz, studying user attitudes rather than autonomous spoken intervention.

LLMs have turned deciding {whether}, {when}, and {what} to contribute into a modeling problem~\cite{deng2023survey,deng2024humancentered,yang2026proactbreakdownawareproactiveagent}. Task agents infer needs from desktop or mobile activity~\cite{lu2024proactive,proagentbench2026,probe2025}, ContextAgent from egocentric sensors~\cite{contextagent2025}, and CollabLLM takes initiative during dialogue~\cite{wu2025collabllm}; all, however, assist a single user rather than compete with humans for the floor. Inner Thoughts scores covert thoughts to select moments to contribute in multi-party text chat~\cite{liu2025inner}, DiscussLLM predicts whether to respond or stay silent each turn~\cite{patel2026discussllm}, and ProMediate selects intervention moments in multi-party negotiation~\cite{liu2025promediate}. These systems operate in turn-based text, where the platform effectively gives the agent its turn, and none grounds intervention in whether the group is missing or misstating a verifiable fact. ProVoice-Bench moves to audio~\cite{provoicebench2026}, but tests reactions to predefined sounds rather than the content of an unfolding discussion. SocialMind displays suggestions on AR glasses~\cite{yang2025socialmind}, while LlamaPIE whispers short cues to one wearer~\cite{chen2025llamapie}. Both provide {private cues} invisible to the group, rather than speaking into the conversation itself.

Our work builds on this prior work, but shared speech introduces constraints that text does not. A chat message, AR overlay, or private whisper can be skimmed, deferred, or ignored while the conversation continues. In contrast, a spoken intervention consumes shared airtime and displaces whoever might speak next. 

\vskip 0.05in\noindent{\bf Turn-Taking, Interruption, and Full-Duplex Spoken Dialogue.}
Conversation analysis shows that turn-taking is tightly organized~\cite{sacks1974simplest,levinson2015timing}, with self-repair preferred over other-repair~\cite{schegloff1977preference}. 
We operationalize it by making the agent a deliberately \emph{dispreferred other-repairer} that waits through the self-repair window. Computational models predict {when a transition is possible}~\cite{ekstedt2020turngpt,ekstedt2022vap,inoue2024realtime}, not whether the agent should speak.  Concurrent work extends turn-taking to multi-party speech under floor competition~\cite{mitra2026adaptiveturntakingrealtimemultiparty}, and full-duplex dyadic models listen and speak at once~\cite{syncllm,defossez2024moshi,avdialog,lin2022duplex}, but both predict {floor availability}, when a role may take a turn, assuming each utterance is addressed to the agent. Whle provide the mechanics for {how} to enter, including VAD-gated delivery and barge-in yielding, but not {whether} to enter in multi-party settings, which requires content-level evidence that the group is missing or misstating something and has not already corrected it.

\vskip 0.05in\noindent{\bf Agents in Meetings and Group Conversations.} Meetings have long been used to study conversation understanding, from multi-party corpora~\cite{carletta2005ami,janin2003icsi} to meeting summarization and assistant benchmarks~\cite{zhong2021qmsum,meetbench2026} and LLM delegates that attend on a user's behalf~\cite{meetingdelegate2025}. These systems either process the meeting afterward or respond to queries; they do not decide on their own that something said was wrong. Other work puts agents into group conversations as moderators, facilitators, or devil's advocates~\cite{kim2020bot,kim2021moderator,mao2024muca,chiang2024devils,alsobay2025bringing,facct2026facilitation}, but in \emph{text}, and mainly to shape participation or opinions. Recent studies examine when agents should participate in groups~\cite{houde2025controlling,readtheroom2026,zhou2026exploring}, highlighting the same tradeoff we face: speaking can disrupt, while silence can miss an opportunity to help. But these systems are formative or user-controlled, rather than autonomous spoken agents, and their open-ended settings provide no ground truth.


\vskip 0.05in\noindent{\bf Grounded Generation and Verifiable Attribution.} Retrieval-augmented generation grounds LLM outputs in sources~\cite{lewis2020rag}, while attributed generation links claims to passages~\cite{gao2023alce,menick2022teaching,xia2025reclaim,zhang2024longcite} and verification pipelines check whether evidence supports those claims~\cite{min2023factscore,vericite2025,guo2022survey}. These methods largely assume a \emph{written} answer to an \emph{explicit} question, checked offline by the person who asked. 


\vskip 0.05in\noindent{\bf Evaluating Proactive Conversational Behavior.}  Proactive-agent benchmarks measure task anticipation~\cite{proagentbench2026,probe2025,proactiveeval2025}, but focus on desktops and datastores rather than conversation. LLM-judged testbeds~\cite{zhou2024sotopia} scale evaluation  while user studies measure helpfulness and intrusiveness~\cite{chen2025llamapie,yang2025socialmind,zargham2022proactivity}, capturing how interventions {felt}. ProMediate is closest to our setting, evaluating intervention timing in multi-party negotiation~\cite{liu2025promediate}, but uses turn-based text with LLM-simulated negotiators. The simulation gives the mediator its turn, so claiming the floor is not an issue, and a negotiation problem setting has no objective answer key for judging an intervention. Our work targets \emph{epistemic} rather than social breakdowns. We seed controlled documents with known facts, predictable gaps, and distractors, turning proactive intervention into a detection task with measurable accuracy and timing error. 




\section{SYSTEM DESIGN}

\begin{figure*}[t!]
  \centering
\includegraphics[width=0.8\textwidth]{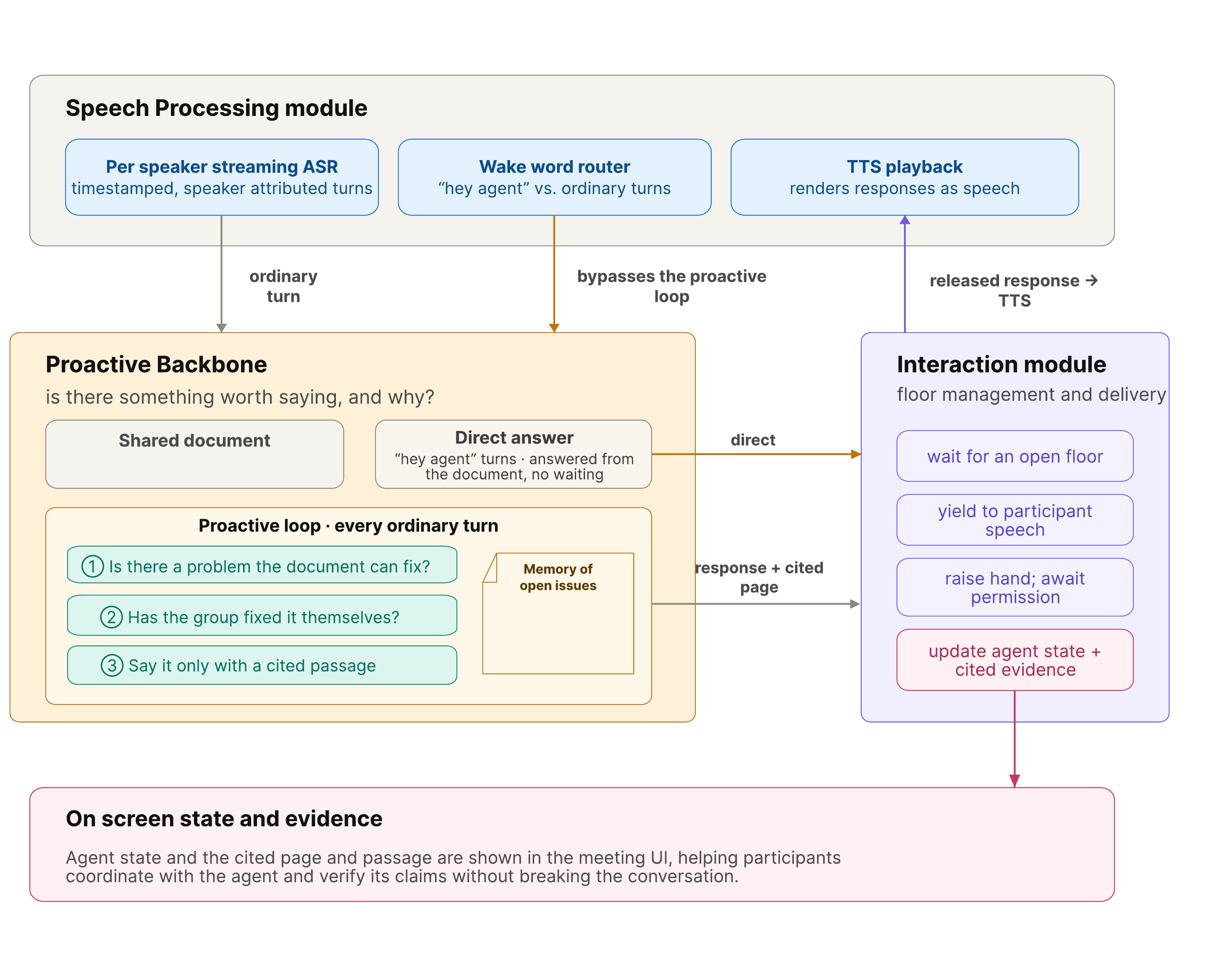}
\vskip -0.25in
  \caption{\textbf{Jarvis system architecture.} It has three  components: speech processing module, proactive backbone and interaction module.}
\vskip -0.15in
\Description{Block diagram of the Jarvis system. The Speech Processing module at the top contains per-speaker streaming ASR, a wake-word router, and TTS playback. Ordinary turns flow to the Proactive Backbone, where the shared document, a memory of open issues, and a three-step proactive loop determine whether to produce a cited response. Direct requests bypass the proactive loop and go to a document-grounded answering path. Both types of responses enter the Interaction module, which waits for an open floor, yields to participant speech, optionally raises a hand and awaits permission, and sends approved responses to TTS. Agent state and cited evidence are also displayed in the meeting interface.}
  \label{fig:system-architecture}
\end{figure*}

As shown in Fig.~\ref{fig:system-architecture}, Jarvis has three modules, each handling a different part of the interaction: who a turn is from, whether there is something worth saying and why, and when and how to say it. The Speech Processing module (\xref{sec:speech}) converts live meeting audio into timestamped, speaker-attributed text for each turn, detects the wake word, and plays the agent's responses back into the meeting. The Proactive Backbone model  (\xref{sec:proactive})  takes the document shared before the meeting and, for each turn from the Speech Processing module, decides whether the agent should contribute and, if so, grounds the response in an exact passage. The Interaction module (\xref{sec:interact})   manages delivery: it waits for an opening rather than talking over the group, yields when someone speaks, and, depending on the mode, either raises a hand for the group to call on or displays the cited page so participants can verify the claim.


\subsection{Speech Processing Module}\label{sec:speech}


It provides three interfaces. First, to the proactive backbone, it sends timestamped, speaker-attributed turns represented as \(\langle \mathrm{speaker}, \mathrm{text}, \mathrm{start\ time}, \mathrm{end\ time} \rangle\), tagged as either \emph{direct} (addressed to the agent) or \emph{stream} (participant-to-participant talk).  Second, to the interaction module, it sends pre-lexical acoustic events such as speech onset, sustained speech, and silence. Finally, it synthesizes approved responses from the proactive backbone agent and plays them back into the meeting as speech. This is  implemented via five sub-components, details of which are in~\xref{sec:speechdetails}. 

\subsection{Proactive Backbone: Deterministic harness of small models}\label{sec:proactive}

\begin{figure*}[t!]
  \centering
  \includegraphics[width=\textwidth]{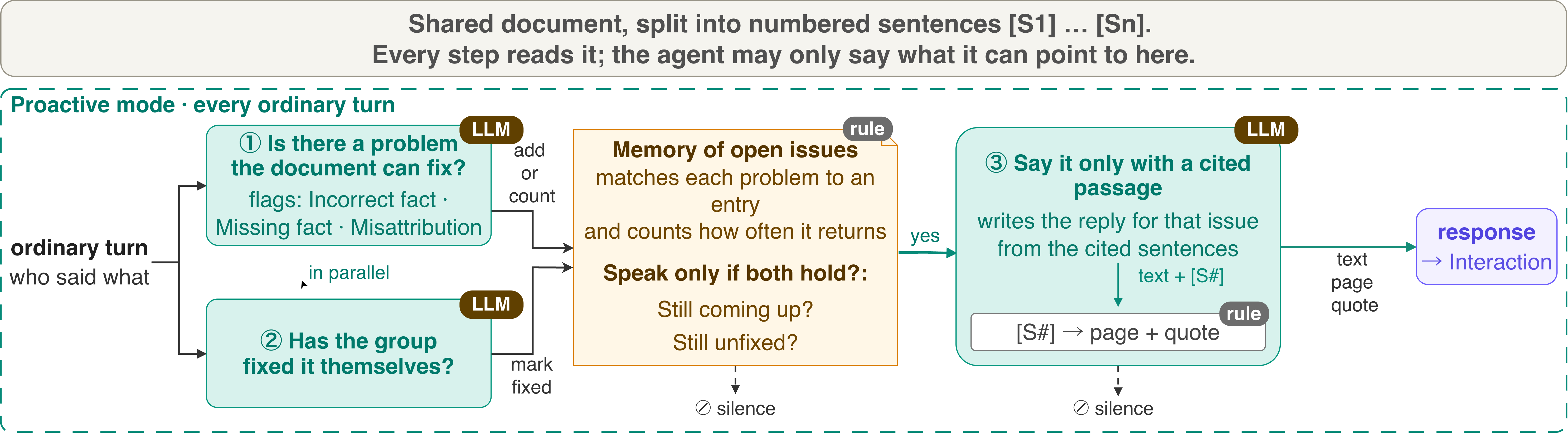}
  \vskip -0.15in
  \caption{\textbf{Our proactive backbone is a deterministic harness of small models.}
For each  turn, the monitor identifies problems that can be addressed using the shared document, while the self-repair check determines whether the group has already resolved them. The two stages run in parallel and update a memory of open issues. A deterministic gate permits an intervention only when an issue persists across turns and remains unresolved. If these conditions hold, Jarvis generates a response supported by cited sentences from the shared document. The cited sentence IDs are resolved to their source page and passage before the response is sent to the Interaction module. Jarvis remains silent when the gate conditions are not met or a valid citation cannot be produced. }
\vskip -0.15in
\Description{Horizontal flow diagram of the Proactive Backbone for ordinary turns. A shared document split into numbered sentences spans the top. Each ordinary turn enters two parallel language-model stages: the monitor flags an incorrect fact, missing fact, or misattribution that the document can address, while the self-repair check determines whether the group has already resolved it. Their outputs add, count, or mark entries as fixed in a memory of open issues. A fixed rule forwards an issue only if it persists across turns and remains unresolved; otherwise, the system stays silent. A final language-model stage generates a response using cited document sentences, and a fixed rule maps the cited sentence IDs to their source page and passage. The response text, page, and passage are then sent to the Interaction module. If a valid citation cannot be produced, the system remains silent.}

  \label{fig:proactive-backbone-architecture}
\end{figure*}


Rather than rely on a single model, we design a deterministic harness  that 
break the task into multiple   steps. Each step uses the same  language model for the part that requires language understanding and deterministic logic to check its output. This prevents any single model decision from triggering an intervention on its own. The backbone has {four}  stages {(a monitor, a self-repair check, a grounded generator, and a direct answerer)} and a deterministic gate (Fig.~\ref{fig:proactive-backbone-architecture}(b)). All stages use the same document, which we split into numbered sentences (\xref{sec:docrep}). Every stage refers to the document only via sentence IDs.  

{Each incoming turn other than backchannels (``yeah,'' ``uh-huh'')  is first appended to the transcript. 
Two} model stages run {in parallel}. The \emph{monitor} (\xref{sec:monitor}) classifies the discussion as either missing or misstating something the document can resolve, or as ordinary conversation that needs no intervention. {The} \emph{self-repair check} (\xref{sec:selfrepair}) looks at the recent conversation and asks whether the participants have already fixed a problem that the monitor flagged {on an earlier turn}. A deterministic  test confirms that  the correct value actually appears in the recent turns before clearing the problem.

The two outputs feed a deterministic \emph{gate} (\xref{sec:gate}), which decides whether Jarvis should speak. The gate keeps a ledger of unresolved problems and intervenes only when a problem persists across turns without being resolved. This gives the group a chance to catch and correct its own mistakes before Jarvis steps in. When the gate fires, \emph{grounded generation} (\xref{sec:generation}) produces the response. It takes the identified problem and the specific document sentences that support the response as input, and generates the response grounded in the document{; a response that cannot be tied to a document sentence is not spoken}. A separate \emph{direct answering} stage ({\xref{sec:direct}}) handles questions explicitly addressed to Jarvis by wake word. These questions bypass the monitor and gate and receive a document-grounded answer. 

\subsubsection{Shared Document Representation}
\label{sec:docrep}
 
 We take the full document text, strip markdown headings and figure apparatus, and split it into sentences on ``\texttt{.!?}'', dropping fragments shorter than five tokens. Each surviving sentence is given an integer ID and listed as a numbered block:
\texttt{[S0]} \textit{sentence 0}, \texttt{[S1]} \textit{sentence 1}, \ldots  ,\texttt{[Sn]} \textit{sentence n}. 
This block is assembled once and prepended to every stage's system message, so all stages index the document by the same IDs.
 
A stage that finds a relevant fact reports the ID of the sentence it came from, and not the text. The server resolves each ID back to its verbatim sentence and drops any ID it cannot resolve. A model can therefore point at the document but does not paraphrase it. This design ensures that the citation is more trustworthy since  the spoken claim is backed by source text the system retrieved, not by text the model produced. We cap the IDs at three per report for the monitor and the self-repair check.  Sharing one representation also has a practical benefit. Because all four system messages begin with the identical document block, the serving engine prefills it once per turn and reuses the cache across stages; only the shorter policy text that follows differs between stages.
 

\subsubsection{Monitor}
\label{sec:monitor}
The monitor performs three {tasks} at each turn: it classifies the discussion's status, extracts a  \emph{claim} naming the point under discussion, and, if  the status is actionable, retrieves the related document sentence IDs. It runs at a low temperature (0.1). Detection, claim identification, and grounding are deliberately fused in one component, which is what lets the monitor {output} a checkable judgment (``this contradicts [S7]'') rather than an opinion (``something seems wrong''). See Table~\ref{tab:app-prompt-monitor} for the full prompt.

\vskip 0.05in   \noindent\textbf{Conversation status taxonomy.}
The monitor assigns each new turn to  one of six statuses.
\squishlist
\item  \emph{Incorrect fact}, where a participant asserts something the document contradicts.
\item \emph{Missing fact}, where the participants need a specific fact the document states but no one has supplied.
\item \emph{Misattribution}, where a participant attributes a claim to the document (``in this study\ldots'', ``they found\ldots'') that the document does not support. 
\item \emph{On track}, accurate and consistent with the text.
\item \emph{Off topic}, concerning logistics, tooling, or side chat that asserts nothing the document could contradict.
\item \emph{Unanswerable from the document}, where  participants are stuck or unsure about a point, the document is about that topic, but  does not actually state the specific detail they need.
\squishend
The first three statuses are actionable, meaning that they name a problem the agent could intervene on, and the ledger tracks them. The other three call for no intervention and open no problem. 

 
 We distinguish \emph{misattribution} from being merely \emph{off topic}. This matters because a mention of another document warrants silence, whereas a false claim about \emph{this} document does not. Similarly, we distinguish a \emph{missing fact} from an \emph{unanswerable} question. This allows the monitor to report a missing fact only when it can point to the exact phrase in the text that answers the specific question. If the document covers the same topic but does not state that particular detail, the status is \emph{unanswerable}, and the agent has nothing grounded to add.   While the monitor uses the recent turns  as context, the status is output for each new utterance. 



\vskip 0.05in\noindent\textbf{Claim identification.} The monitor also outputs a \emph{claim}: a short noun phrase naming the point under discussion, independent of the answer or participants' wording (e.g., ``the dataset size used for training,'' not ``recruiting 24 participants through university mailing lists''). This normalizes different surface forms (``24 seems small,'' ``how many did they recruit?'') to one claim, and the monitor reuses it as long as the discussion stays on that point, regardless of status: if the group first asks about $X$ and then says something wrong about $X$, both turns map to $X$. This canonical form is what lets the backbone track a problem across turns and decide when it has persisted (\S\ref{sec:gate}). The monitor also returns the \emph{evidence}, the participant's words that triggered the status.

\vskip 0.05in\noindent\textbf{Retrieving sentence ID.} Finally, if the status is actionable, the monitor returns the answer along with its source: the actual text/value pulled from the document that the group got wrong, copied verbatim, and the IDs of the sentences it came from (at most three). Where the value sits in a table or figure and no sentence can be cited, it returns the value with no IDs rather than a fabricated one.  These IDs  serve as a fallback for the generation stage (\xref{sec:generation}) if its own citation cannot be resolved; if neither yields a resolvable sentence, the correction is not spoken. 


\subsubsection{Self-Repair Check}
\label{sec:selfrepair}  For each problem, this language model  asks: ``Has this problem since been fixed by the participants themselves?'' It runs at a low temperature of 0.1.


Given the document, the list of open claims  and the same recent transcript the monitor reads, the self-repair model checks each open claim in turn. It first locates the sentence stating the correct fact. It then applies the following rule: a claim is marked resolved  if the transcript has \emph{stated} that correct fact.  Our prompt  is explicit about what does {not} count. For example,  repeating the same wrong value, merely mentioning the topic again, or a participant admitting they cannot recall do not count as resolution. We only consider the recent transcript, and the model outputs  the claim IDs it judges resolved, citing the grounding sentence IDs for each.




\noindent\textbf{Filtering the model's checks.} Each proposed resolution passes through a deterministic guard that rechecks the transcript before closing a claim. The guard closes a claim only if the correct value appears in the last three turns, with at least 60\% of its content tokens present (question words removed, numbers matching exactly). It does not accept a question, nor an utterance that reports failure to resolve (``I can't recall''), and the closing utterance must add words absent from the turns that first raised the claim. A misattribution has no document value to verify; there the guard instead accepts a retraction (``that's not from this paper''), and for any other valueless claim it lets the model's judgment stand. The model and the guard are complementary: the model captures meaning a string match cannot, such as that ``two dozen'' answers ``how many participants,'' while the guard checks that the value was actually stated, and stated recently. The guard sees only the last three turns; the model reads the full window.
 

\subsubsection{Claim Ledger and Gate}
\label{sec:gate}

Neither the monitor nor the self-repair check decides whether Jarvis speaks; that decision is made by our {gate}, which maintains a {claim ledger}.

Each turn, the monitor may report a claim, and we check whether it matches one already open. We compare claims by sentence embedding (\textsc{all-MiniLM-L6-v2}~\cite{reimers-gurevych-2019-sentence}) and treat them as the same problem when cosine similarity is at least $\sigma$. When an actionable claim matches an open one, we increment its \emph{persistence count}, the number of consecutive detections without an intervening turn that reports the point as non-actionable. A turn that does not mention the point leaves the count unchanged. One that reports it with a non-actionable status resets it to zero. A claim matching nothing open is added with count one, and a closed claim asserted again is reopened from one.


The gate fires when four conditions hold: the persistence count reaches a threshold $\tau$; the self-repair check has not closed the claim; the claim is not in a cooldown; and we have not already spoken it. A cooldown is $\rho$ turns of silence after an intervention or a participant resolution, during which the claim cannot fire. A fact counts as already spoken only when the same question recurs soon after it was answered; a repeated wrong assertion is corrected again. Higher $\tau$ makes Jarvis more patient, lower $\tau$ makes it intervene sooner.


Because a problem flagged once does not fire immediately, participants have $\tau{-}1$ turns to correct it themselves. The self-repair check runs each turn and its resolutions are applied before the gate tests the count. If the group fixes the problem first, the claim is closed and the gate does not fire, so Jarvis speaks only when a problem persists unresolved.


 When the gate fires, we pass the claim to grounded generation (\S\ref{sec:generation}), which may abstain if the document supports no answer, or if the generated text cites no valid sentence. We then keep the claim open for later, up to $\alpha$ such rollbacks. A rolled-back claim fires again on its next detection and starts no cooldown; after the $\alpha$-th rollback we treat it as spoken, so a claim the document cannot ground stops firing. Finally, the ledger forgets: a closed claim is dropped after $W_r$ turns, and an open one after $W$ turns without a mention.


{In the deployed configuration $\tau=2$, $\rho=6$, $\alpha=2$, $\sigma=0.6$, $\delta=0.5$, and $W=W_s=W_u=10$, $W_r=3$ turns.}

\subsubsection{Grounded Generation}
\label{sec:generation}

 This  runs only on a fired claim, at a higher temperature ($0.7$), and streams its output so speech can begin before the full response is ready.


The intent follows from the monitor's status, and there are three. For \emph{correct}, the model verifies the claim against the cited sentences and, if it is wrong, states the correction in one or two sentences (at most 25 words). If the sentences neither support nor contradict it, the claim is treated as a misattribution. For \emph{missing}, the model answers completely, with no length limit. For \emph{misattribution}, it says in one short sentence (at most 20 words) that the document does not state the claim. The correct and missing intents must cite sentences by ID, ordered most-substantive first. The first cited sentence sets the page the listener is pointed to. If the cited IDs cannot be resolved, the response falls back to the monitor's IDs, and if those are empty too it is discarded (\S\ref{sec:gate}).

The model outputs spoken \texttt{text} without IDs, plus a \texttt{topic}: a two-to-four-word lowercase noun phrase naming what the answer is about, to orient a listener who did not ask. We assemble the utterance as ``About \{topic\}, \{text\} You'll find this on \{Page N\}; search for `\{keyword\}' to confirm,'' and a client may reduce this to the page alone or turn it off. Before delivery, the utterance is compared against corrections from the last $W_u$ turns and dropped if it is a near-duplicate (embedding similarity $\ge\delta$) of one on the same claim. The delivered utterance is appended to the transcript, so the self-repair check can later recognize Jarvis's own correction as a resolution.

\subsection{Interaction Module}\label{sec:interact}



This module explores three methods, described below, and is implemented as an event-driven state machine driven by the turn, wake-word, and voice-activity events from the speech module (see \xref{sec:direct-entry} for how it handles direct questions).



\subsubsection{Proactive Entry, Barge-in, and Recovery}
\label{sec:floor-timing}
 
The module uses silence as its cue for an opening, but silence is ambiguous: a pause may mean an open floor or a speaker planning their next clause. We therefore do not play an intervention as soon as it is generated. We queue it and release it only when three conditions hold: VAD has detected a long silence, no participant is speaking, and input energy has stayed low for a minimum interval. Together these distinguish a real opening from a within-utterance pause. If the backbone generates a newer intervention first, it replaces the queued one, so Jarvis responds to the current state of the discussion.

Once speaking, Jarvis gives the floor back as soon as a participant takes it, triggered only by sustained speech so brief backchannels do not interrupt it. What happens to the interrupted response depends on how much was delivered: interrupted almost immediately, Jarvis replays it in full at the next opening. Interrupted after a few seconds, it resumes from the start of the truncated sentence with ``To finish my response, \ldots''. If enough was already spoken, it discards the rest, since an interruption can mean either disinterest or that the point already landed. 


The backbone may send an intervention as complete text or as sequence-numbered streaming deltas. We discard duplicates and outputs from superseded generations on arrival. This barge-in and recovery logic governs all Jarvis playback, including direct answers. In our implementation, the playback gate opens after $2.0$~s of VAD silence with all participant VADs silent and input RMS below $0.004$ for $350$~ms; $0.5$~s of continuous participant speech stops Jarvis. Recovery follows delivered duration: under $2$~s replays in full, $2$-$5$~s resumes at the interrupted sentence, and beyond $5$~s discards the rest.


\subsubsection{Asking for Permission: Hand-Raising}
In \emph{hand-raise} mode, Jarvis holds its response and raises its hand, speaking only after a participant gives it the floor, either by using the wake word or an explicit phrase such as ``go ahead, agent'' without immediately continuing. If no one calls on it within 60~s, the response is discarded and the hand lowered, so stale information is not delivered.

We detect permission with end-anchored regular expressions for phrases such as ``go ahead (agent),'' ``agent, (please) go on/speak/answer/jump in,'' and ``(agent,) you can speak/talk/answer,'' while rejecting denials such as \emph{not yet}, \emph{hold on}, \emph{wait}, and \emph{stop}. Because ASR often drops ``agent,'' we also accept bare ``go ahead'' or ``you can speak'' while the hand is raised. ASR can also split a phrase across sub-phrases (``agent'' then ``go ahead''). We keep partial sub-phrases containing ``agent'' or ``go ahead,'' append adjacent ones from the same speaker within 1.5~s and a 3~s total span, and match the concatenated text. After detecting permission, Jarvis waits 400~ms and, if no new speech begins, releases the response.

\subsubsection{Visually communicating agent state and evidence}
\label{sec:visibility}


Participants need to know whether Jarvis heard them, is thinking, or is about to speak; without this, a missed wake word leaves them unsure whether to repeat, and an unexpected response feels like an interruption. We show Jarvis's state (\emph{idle}, \emph{listening}, \emph{thinking}, \emph{speaking}) through three channels: a status glyph next to its name (an ear, thought bubble, or speaking head), a state-colored circle on its video tile, and a brief earcon on wake-word detection. In hand-raise mode, the raised hand also signals it has something to say.

Participants may not trust the agent's claims without evidence, so we show the source as well as read the claim aloud. Each intervention carries three pieces of grounding metadata: a page, keywords, and a source passage. While Jarvis speaks, the system sends this to the shared PDF viewer, which scrolls to the cited page and marks the passage with an arrow, letting participants check the claim against the source without leaving the conversation. The \emph{listening} and \emph{thinking} name suffixes time out after $30$~s and $15$~s, state updates are debounced by $250$~ms, and the video tile renders as a $1280\times720$ source at $15$~fps. If grounding is unavailable, Jarvis instead mentions ``page $N$'' in the spoken response.




\section{Synthetic Dataset Generation}
\label{sec:dataset}

We introduce \texttt{CHI-180-proactive}, built from 180 CHI papers (2008-2025), ten from each of the 18 subcommittees. For each paper we synthesize a discussion between two researchers who have read it and plant 24 problems, which we call \emph{events}, giving 4{,}320 events over 50{,}400 turns. We describe the events below, then the three-stage pipeline that generates the data (details in~\xref{app:dataset}).

\begin{table*}[t!]
\caption{The five planted event types and their ideal actions.}
\Description{Table listing five event types (gap, wrong, self-correction, resolved question, and unanswerable), followed by their definitions, classes, cues, ideal actions, and counts per paper.}
\label{tab:events}
\small
\vskip -0.15in
\setlength{\tabcolsep}{3pt}

\begin{tabular}{@{}
p{0.14\linewidth}
p{0.30\linewidth}
p{0.12\linewidth}
p{0.07\linewidth}
p{0.19\linewidth}
c@{}}
\rowcolor{black!7}
\textbf{Event}
& \textbf{Definition}
& \textbf{Class}
& \textbf{Cue}
& \textbf{Ideal action}
& \textbf{Per paper} \\

\rowcolor{JarvisGapBlue!16}
\cellcolor{JarvisGapBlue!38}\raggedright\textbf{Gap}
& A question answered by the document is asked, no participant answers it, and the discussion continues
& Must speak
& Question
& Answer with a citation
& 10 \\

\rowcolor{JarvisWrongRed!16}
\cellcolor{JarvisWrongRed!38}\raggedright\textbf{Wrong}
& A claim contradicting the document is asserted, and the other speaker adopts it over the next two turns
& Must speak
& Claim
& Correct after a grace period
& 4 \\

\rowcolor{JarvisSelfCorrectionYellow!16}
\cellcolor{JarvisSelfCorrectionYellow!38}\raggedright\textbf{Self-correction}
& A false claim that the participants repair themselves
& Must stay silent
& Claim
& Stay silent (already resolved)
& 4 \\

\rowcolor{JarvisResolvedquestionGreen!16}
\cellcolor{JarvisResolvedquestionGreen!38}\raggedright
\textbf{Resolved question}
& A question answered by the document is asked, and the other speaker states the document's answer in full
& Must stay silent
& Question
& Stay silent (nothing to add)
& 4 \\

\rowcolor{JarvisUnanswerableGray!16}
\cellcolor{JarvisUnanswerableGray!38}\raggedright
\textbf{Unanswerable}
& A question the document does not answer
& Must not fabricate
& Question
& Stay silent, explicitly abstain, or offer only document-grounded related information
& 2 \\

\end{tabular}
\end{table*}

\subsection{Events and Ideal Actions}
\label{sec:dataset-events}

Each event is a short passage containing one thing the agent could respond to: a question asked and left open, or a false claim asserted. Every event records the turn where the trigger appears, the \emph{ideal action} (which for some events is to stay silent), and a \emph{window of usefulness}, the span of turns during which that action still helps, since a correction that arrives after the group has moved on no longer counts. There are five event types (Table~\ref{tab:events}), split by whether the agent should speak or stay silent. Everything else is ordinary conversation: of the roughly 280 turns per paper, about 256 carry no event, and silence is correct on all of them.


Silent events serve as controls. Each speak-event has a silent counterpart with the \emph{same trigger}: a \emph{gap} (an open question the document answers) pairs with a \emph{resolved question} (the same question, answered correctly by a participant), and a \emph{wrong} (a false claim left to stand) pairs with a \emph{self-correction} (a false claim the participants repair). In both silent cases the trigger is present, only the group's own resolution differs. An ordinary silent turn cannot test this, but a matched pair can: it asks whether the agent distinguishes a resolved trigger from an unresolved one, and speaking on a resolved one is an unnecessary interruption rather than a caught error. This lets us measure detection failures and false interruptions separately.



We create two ``wrong'' error types. A \emph{number swap} replaces one numeric value with a nearby value of the same unit and magnitude, testing fine-grained grounding. A \emph{cross-paper misattribution} states a distinctive fact from a different CHI paper as if it belonged to this one, testing whether the agent is bound to {this} document: an agent that retrieves from the wider literature will find something similar elsewhere and wrongly accept it.

\subsection{Generation Pipeline}
\label{sec:dataset-pipeline}
 
 

We build each paper's data in three stages (details in~\xref{app:dataset}).
 
\noindent\textbf{Stage 1: Facts and questions.} Using Qwen3-235B-A22B-Instruct-2507~\cite{qwen3}, we extract every fact the paper states, anchored to its page and passage, and synthesize twelve question-answer items per paper (four extractive, four abstractive, two boolean, two unanswerable), each with a gold answer, acceptable variants, and its source evidence. The count and answer types follow QASPER~\cite{dasigi2021qasper}. These items become the gap and unanswerable events, and also benchmark the agent's question answering (\S\ref{sec:evaluation}), so the two evaluations run over identical questions.

\begin{table}[t!]
\vskip 0.1in
\caption{Generator screening. Each value is the Jensen--Shannon divergence between the distribution of the judge's scores over 38 synthetic eight-turn windows and over 38 windows of ArgSciChat: 0 means the two score distributions are identical, and the value grows the less they overlap. For scale, two halves of ArgSciChat itself differ from each other by 0.14 to 0.19, depending on the axis. Q235 = Qwen3-235B-A22B, GPT = GPT-OSS-120B, Q27 = Qwen3.8-27B, DS = DeepSeek-V4-Flash-0731. Condition 2, in bold, has the smallest mean and is the one we use.}
\Description{Table of 16 generator conditions with Jensen-Shannon divergence values on naturalness, coherence, engagingness, and groundedness, and how many of the four axes each condition cleared.}
\vskip -0.15in
\label{tab:generator}
\small
\begin{tabular}[]{@{}rlrrrrc@{}}
\toprule
\# & Asker $\times$ explainer & Natural. & Coher. & Engag. & Ground. & Mean \\
\midrule
1 & Q235 $\times$ Q235 & 0.167 & 0.067 & 0.005 & 0.104 & 0.086 \\
\textbf{2} & \textbf{GPT $\times$ GPT} & \textbf{0.142} & \textbf{0.035} & \textbf{0.011} & \textbf{0.064} & \textbf{0.063} \\
3 & Q27 $\times$ Q27 & 0.029 & 0.045 & 0.195 & 0.187 & 0.114 \\
4 & DS $\times$ DS & 0.055 & 0.083 & 0.151 & 0.013 & 0.076 \\
5 & Q235 $\times$ GPT & 0.272 & 0.107 & 0.013 & 0.047 & 0.110 \\
6 & GPT $\times$ Q235 & 0.156 & 0.123 & 0.012 & 0.074 & 0.091 \\
7 & Q235 $\times$ Q27 & 0.075 & 0.076 & 0.261 & 0.103 & 0.129 \\
8 & Q27 $\times$ Q235 & 0.070 & 0.020 & 0.171 & 0.276 & 0.134 \\
9 & Q235 $\times$ DS & 0.215 & 0.076 & 0.081 & 0.027 & 0.100 \\
10 & DS $\times$ Q235 & 0.090 & 0.057 & 0.102 & 0.064 & 0.078 \\
11 & Q27 $\times$ GPT & 0.090 & 0.048 & 0.097 & 0.136 & 0.093 \\
12 & GPT $\times$ Q27 & 0.171 & 0.090 & 0.063 & 0.091 & 0.104 \\
13 & Q27 $\times$ DS & 0.089 & 0.126 & 0.154 & 0.035 & 0.101 \\
14 & DS $\times$ Q27 & 0.037 & 0.038 & 0.182 & 0.089 & 0.087 \\
15 & GPT $\times$ DS & 0.115 & 0.109 & 0.030 & 0.046 & 0.075 \\
16 & DS $\times$ GPT & 0.046 & 0.077 & 0.156 & 0.016 & 0.074 \\
\bottomrule
\end{tabular}
\vskip -0.15in
\end{table}

\noindent\textbf{Stage 2: Planting errors.} We plant eight false claims per paper: four that nobody corrects (\emph{wrong}) and four that the participants repair (\emph{self-correction}). The two share a claim turn and differ in what follows. {In each self-correction, the other speaker gives a short reply on the turn after the false claim, and on the following turn the speaker who made the claim repairs it; the agent must therefore hold back for two turns rather than react to the claim.} Every claim passes deterministic checks that it is genuinely false, not decidable from the conversation alone, non-colliding with the twelve questions, and non-leaking (\xref{app:dataset}).



\noindent\textbf{Stage 3: Conversation synthesis.} We fix a plan before generating text: events ten turns apart, the first no earlier than turn 15, split across a 220-turn \emph{main} conversation with 20 events and a 60-turn \emph{resolved-question} conversation with the other 4. Keeping resolved-question events in a separate conversation ensures the agent's silence there reflects its reading of the group, not its own repetition suppression, and gives us a stretch with no planted problem, where we measure how often the agent speaks when nothing needs saying. The two speakers are researchers deciding whether the paper's evidence justifies a follow-up study, one questioning the evidence and one explaining the methods; non-scripted turns come from an Inner-Thoughts-style engine~\cite{liu2025inner} in which each speaker forms a private thought, rates its motivation, and the more motivated one takes an open floor.



We screened four open-weight models across both speaker roles (16 assignments), scored short windows with a condition-blind, G-Eval-style judge~\cite{liu2023geval} on naturalness, coherence, engagingness, and groundedness, and compared the score distributions against real research conversations by Jensen--Shannon divergence following Mirage~\cite{devanathan2026mirage}, where 0 means identical distributions (Table~\ref{tab:generator}). We use this screening to confirm that the candidate generators produce conversations within the range of variation seen in real research dialogue, rather than to rank them: fifteen of the sixteen conditions fall below the 0.14--0.19 split-half divergence of ArgSciChat itself, so the screening does not reliably discriminate among them at this sample size. We adopt GPT-OSS-120B in both roles as one adequate configuration.

A paper enters CHI-180-proactive only after we verify that all 24 events are present with their oracles, that both conversations are complete at 220 and 60 turns, and that all identifiers agree. Each oracle stores the ideal action (answer, correct, abstain, or stay silent), the expected answer, and the window of usefulness.



\section{Evaluation}
\label{sec:evaluation}
\subsection{Benchmark results}

\subsubsection{Benchmarking  our small models}

\providecommand{\tofill}[1]{{\color{red}[#1]}}
\begin{table*}[t]
\caption{Document-grounded QA accuracy, citation validity, and latency for four models on three datasets.  TTFT is time to first token and Total is time to the last token, both as reported by the inference server. Bold marks the best value per dataset.}
\vskip -0.15in
\label{tab:results}
\centering
\small
\begin{tabular}{@{}lccccc@{}}
\toprule
& \multicolumn{2}{c}{Accuracy} & & \multicolumn{2}{c}{Latency (ms, p50)} \\
\cmidrule(lr){2-3} \cmidrule(l){5-6}
      Model & Strict & Strict+Partial & Citation & TTFT & Total \\
\midrule
\multicolumn{6}{@{}l}{\textit{QASPER} (NLP papers,  $n=682$, 28k-token documents)} \\
      Qwen3.6-35B (deployed) & \textbf{.864} & \textbf{.974} & .919 & 368 & 1,192 \\
      Qwen3.5-35B & .826 & .968 & \textbf{.927} & 361 & 1,376 \\
      Qwen3.5-9B  & .789 & .949 & .877 & 335 & 1,440 \\
      Qwen3.5-4B  & .773 & .946 & .886 & \textbf{318} & \textbf{881} \\
\addlinespace
\multicolumn{6}{@{}l}{\textit{COVID-QA} (biomedical papers,  $n=1{,}829$, 28k-token documents)} \\
      Qwen3.6-35B (deployed) & \textbf{.917} & .979 & .929 & 378 & 1,101 \\
      Qwen3.5-35B & .915 & \textbf{.984} & \textbf{.934} & 377 & 1,277 \\
      Qwen3.5-9B  & .908 & .973 & .920 & 347 & 1,312 \\
      Qwen3.5-4B  & .864 & .965 & .908 & \textbf{325} & \textbf{874} \\
\addlinespace
\multicolumn{6}{@{}l}{\textit{CHI-180} (CHI papers,  $n=1{,}545$, 48k-token documents)} \\
      Qwen3.6-35B (deployed) & .926 & .977 & \textbf{.913} & 443 & 1,189 \\
      Qwen3.5-35B & \textbf{.930} & \textbf{.980} & .910 & 490 & 1,444 \\
      Qwen3.5-9B  & .897 & .966 & .892 & 396 & 1,398 \\
      Qwen3.5-4B  & .874 & .955 & .870 & \textbf{368} & \textbf{889} \\
\bottomrule
\end{tabular}
\vskip -0.15in
\end{table*}

 
We evaluate the direct-answer path, where a participant asks Jarvis a question and it answers from the document and cites the sentences that support the answer. We hold everything fixed except the answering model: the prompts, the guided decoding, and the sampling temperature (0.7) are the same. The evaluation differs from deployment in three ways: each question is injected as a direct turn, bypassing the wake word; the conversational context is empty, since the questions are independent; and we score the answer text and the cited sentence IDs, before the spoken wrapper (``Page~$N$; search for \ldots'') is added. All four models are served in turn on one node (2$\times$L40, vLLM~\cite{kwon2023efficient}~0.26).

\vskip 0.05in\noindent{\bf Models.}
We compare the deployed
\texttt{Qwen3.6-35B-A3B}~\cite{qwen36_35b_a3b}
with three official Qwen3.5 checkpoints~\cite{qwen35}:
\texttt{Qwen3.5-35B-A3B},
\texttt{Qwen3.5-9B}, and
\texttt{Qwen3.5-4B}.
The two 35B models are mixture-of-experts models with 3B activated
parameters.

\vskip 0.05in\noindent{\bf Datasets.}
We use three document-grounded QA sets:
\begin{itemize}
\item \textbf{CHI-180 QA}: The twelve questions per paper from \S\ref{sec:dataset}, the same items reused as gap and unanswerable events in \S\ref{sec:dataset-pipeline}. We ran 2,150 of the 2,160 items (10 had gold evidence that could not be mapped to a sentence) and excluded the 360 unanswerable items from accuracy, as they have no gold answer. Documents were long, at 48k tokens.
\item \textbf{QASPER}~\cite{dasigi2021qasper}: NLP papers, where questions were written by readers who had seen only the title and abstract and answered by others who read the full text. 
\item \textbf{COVID-QA}~\cite{moller2020covidqa}: an extractive set over COVID-19 papers, where annotators marked a span and wrote a question for it. It was released independently of our pipeline.
\end{itemize}
The CHI-180 questions were generated by a Qwen model, so they may favor the Qwen family; QASPER and COVID-QA are not subject to this, which is why we include them.

\vskip 0.05in\noindent{\bf Metrics.} We evaluate three key metrics.
\begin{itemize}
\item \textbf{Accuracy.} An LLM judge (Claude Sonnet~5; rubric in Table~\ref{tab:app-judge-qa}) labels each answer \textsc{Correct}, \textsc{Partial}, \textsc{Wrong}, or \textsc{Gold\_broken}; the last category is when the question or its reference answer is itself broken. We use a judge rather than token overlap because the system answers in full sentences while our gold answers are short, so overlap F1 is low even for correct answers. \emph{Strict} is defined as the fraction   of \textsc{Correct} answers; \emph{Strict+Partial} also counts \textsc{Partial} answers that omit a required element. \textsc{Gold\_broken} items, where the reference itself is defective, leave the denominator.
\item \textbf{Citation.} Among \textsc{Correct} answers, the fraction where the cited sentences overlap the annotated evidence. This is a lower bound, since papers often states the same fact in several places and only one is annotated.
\item \textbf{Latency.} Median time to first token (TTFT) and to the last token (Total) on the answer-generation call. The document prefix is cached first, as is typical in practical deployments.
\end{itemize}
We compare models only on the questions that all four answered and that were judged, dropping any question marked \textsc{Gold\_broken} for any model. This matters because each model refuses a different set of questions, and scoring each model on its own answers would reward the one that refuses more of the hard ones. 

\vskip 0.05in\noindent{\bf Results.}
Table~\ref{tab:results} shows that the two 35B models are close everywhere. The drop begins below 9B, and this also  depends on the dataset. On QASPER and CHI-180, which have multi-part answers and longer documents, the 9B model is  behind and the 4B model further still. Much of the loss is partial rather than wrong: on QASPER only 2.6\% of the deployed model's answers are \textsc{Wrong} while 11.0\% are \textsc{Partial}, because questions like ``which datasets were used'' have several-part answers and giving only some parts scores as \textsc{Partial}. Under Strict+Partial scoring the four models spread by at most .028.

Among correct answers, cited sentences overlap the annotated evidence .91-.93 of the time for the 35B models. These are however lower bounds. When we manually examined the 60 citation misses among the deployed model's 643 \textsc{Correct} QASPER answers, most were near the gold sentence or cited a different sentence that in fact supports the answer: counting those as valid raises the match rate from .907 to .980. So the true citation validity is higher than the table shows, and the gap between models is smaller than it looks.
 
 The 4B model has the shortest TTFT on all three sets  and the shortest total time. The surprise is in total latency  for the mid size 9B model, which is slower end-to-end than the 35B models. This is because the 9B model is dense and outputs 69-71 tokens/s, whereas the mixture of expert (MoE) 35B models has only 3B active parameters and output  84--101 tokens/s. Longer documents cost prefill time: on CHI-180's 48k-token documents, TTFT is 40-130~ms higher for every model than on the 28k-token sets.

\begin{figure*}[t!]
  \centering
  \begin{minipage}[t]{0.48\linewidth}
    \centering
    \includegraphics[width=0.8\linewidth]{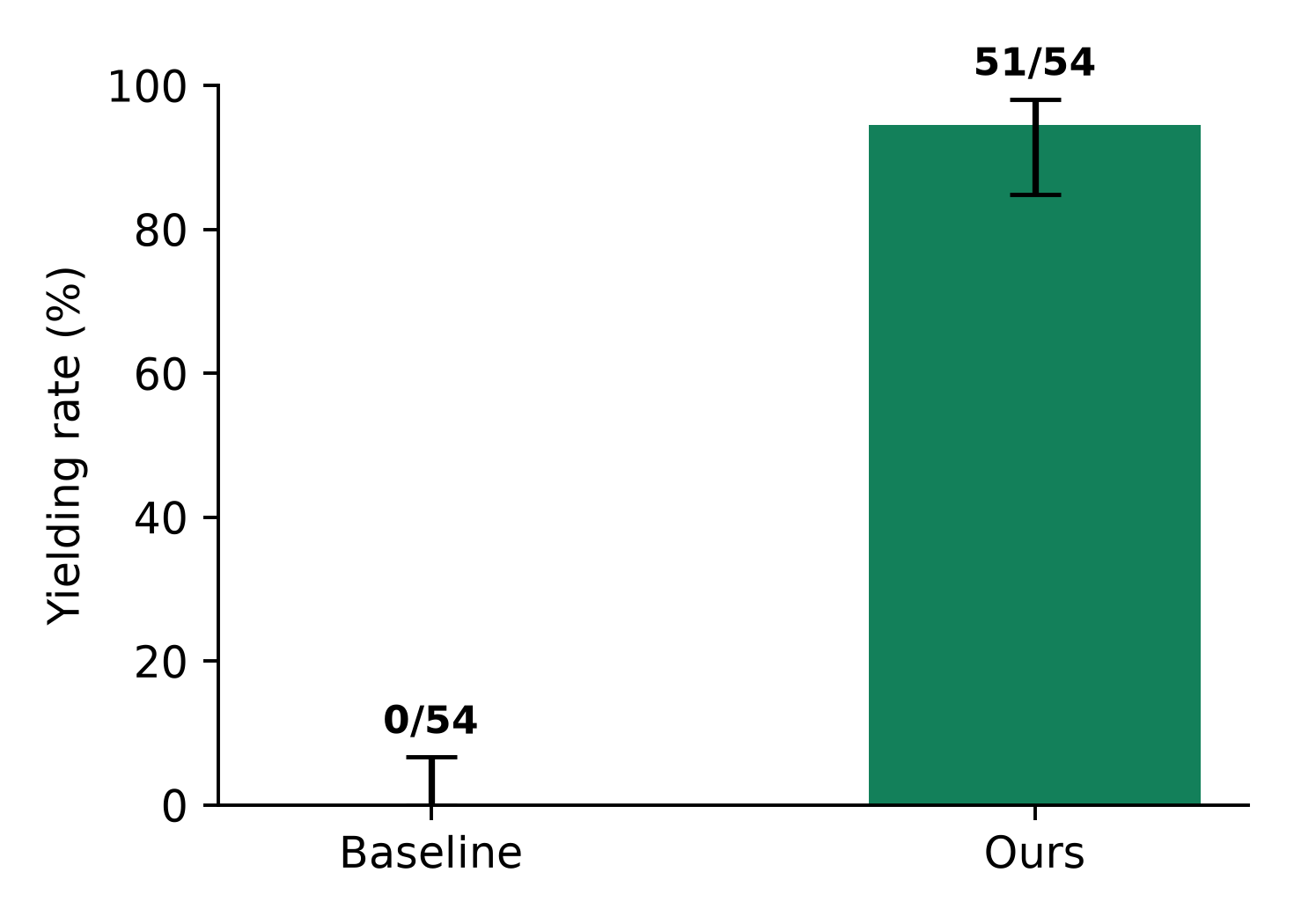}
    \par\small (a) Yielding rate
  \end{minipage}\hfill
  \begin{minipage}[t]{0.48\linewidth}
    \centering
    \includegraphics[width=0.8\linewidth]{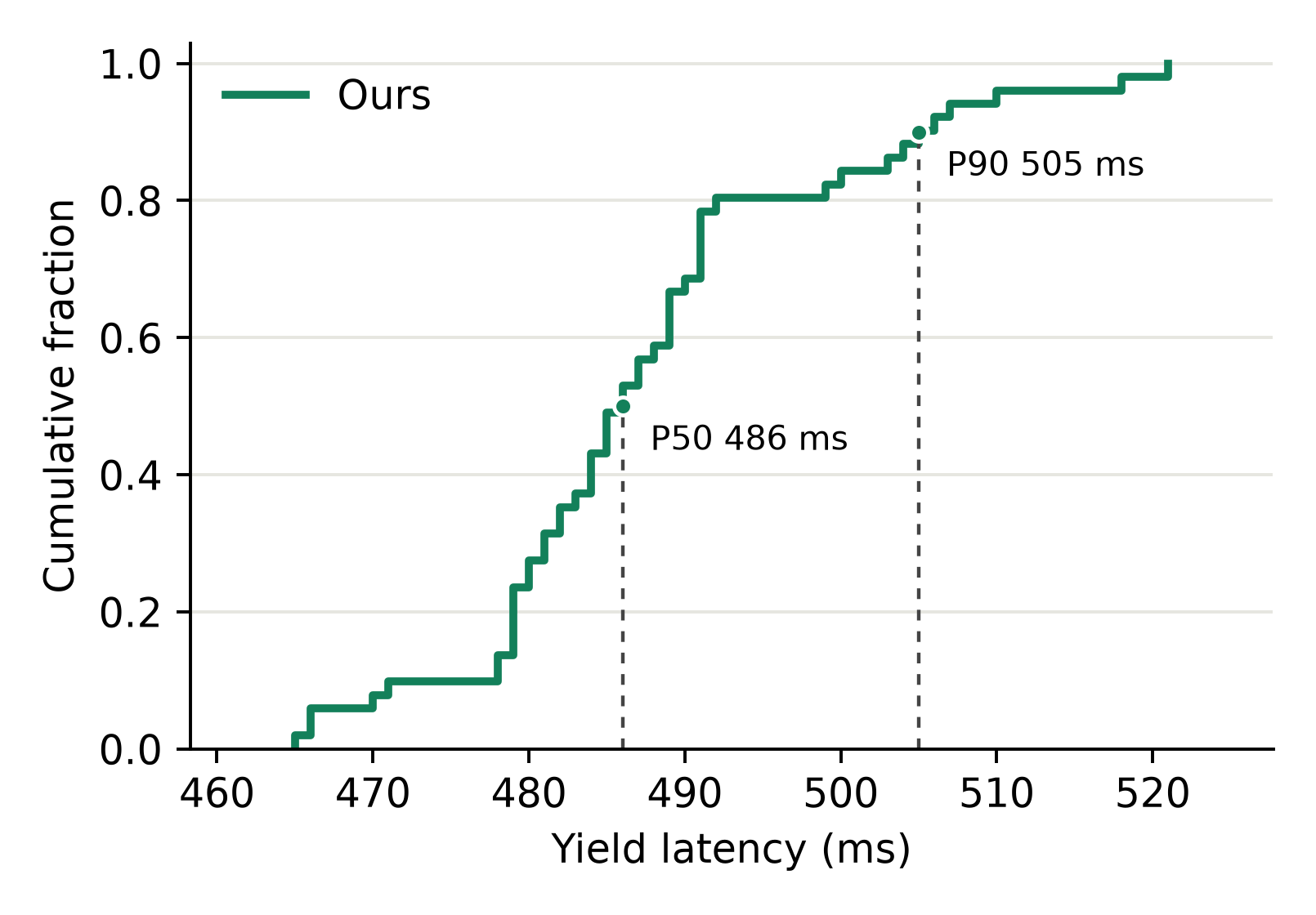}
    \par\small (b) Yielding latency, successful trials only
  \end{minipage}
  \vskip -0.15in
  \caption{\textbf{Barge-in yielding.}
  (a) Yielding rate with 95\% Wilson CIs (54 trials per condition). (b) Empirical CDF of yield latency over the 51 trials in which Ours yielded; the axis is normalized by those 51, not all 54. Baseline never yielded, so it has no CDF. Latency spans VAD detection to the bridge halting audio output, excluding buffering on both sides.
}
\vskip -0.15in
  \label{fig:barge-in-eval}
\end{figure*}

\subsubsection{Benchmarking Agent Yielding Behavior When a Person Speaks}
\label{sec:barge-in-eval}
A proactive agent has to give up the floor the moment a participant talks over it. We evaluate whether Jarvis does, and how quickly its audio stops when humans  start talking. The same mechanism yields for both proactive interventions and answers to direct questions, so we elicit agent speech with direct questions, which are easier to control.

We drew 54 questions from CHI-180-proactive  (three questions from each of 18 papers, one paper per subcommittee). Each question was run once with barge-in yielding \emph{enabled} and once \emph{disabled}, giving 108 trials. In each trial a human participant starts speaking while the AI agent is speaking.  The two conditions differed only in whether Jarvis stopped speaking when it detected sustained human speech. In every trial an experimenter asked the question and then, while Jarvis was answering, interrupted with sentences like ``Sorry, let me add something here. I have another question.''

We report yield  rate per condition with 95\% Wilson confidence intervals. Fig.~\ref{fig:barge-in-eval}(a) shows that with yielding enabled, Jarvis stopped speaking in 51 of 54 trials (94.4\%, 95\% CI 84.9--98.1\%). With it disabled, it stopped in 0 of 54, running to the end of its answer every time (0.0\%, 95\% CI 0.0--6.6\%). 

On the 51 trials that stopped, the median yield latency was  486~ms and latency had a range of 465--521~ms (Fig.~\ref{fig:barge-in-eval}(b)). This is well under the roughly 500~ms-1~s gap that separates turns in human conversation, so from a participant's side the agent yields within about the time it takes one person to start talking after another stops. The three failures were cases where  the interruption did not register as sustained speech.

\subsection{Evaluating Jarvis on synthetic conversations}

\label{sec:eval-proactive}

\begin{figure*}[t!]
\centering
\includegraphics[width=0.9\linewidth]{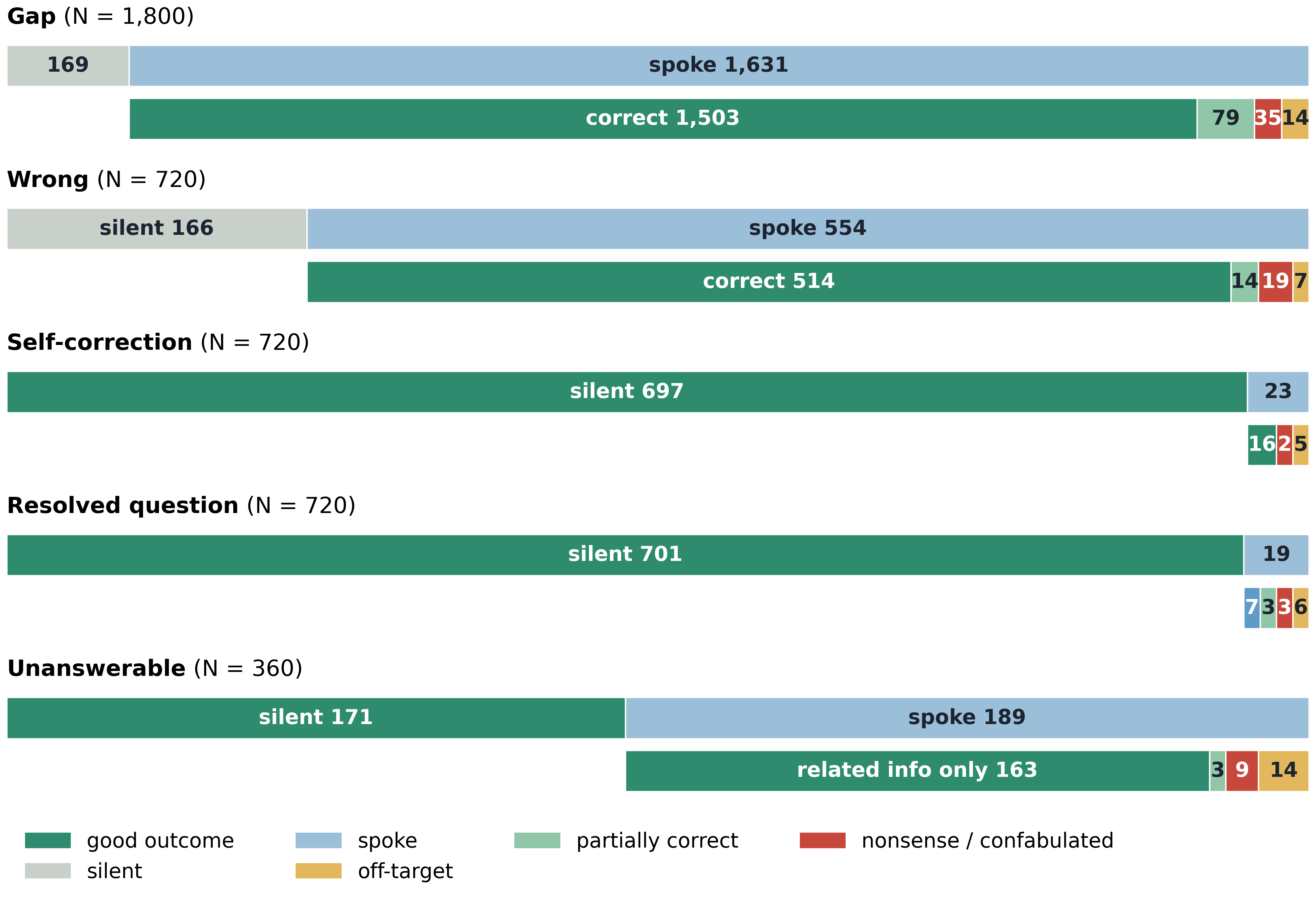}
\vskip -0.15in
\caption{Outcome flow by proactive behavior  across the conversations in the \texttt{CHI-180-proactive} dataset.}
\Description{Five stacked-bar cascades, one for each situation, showing outcomes in patient mode. For each situation, the first bar splits events into silent and spoke, the second splits spoken events into addressed and off-target, and the third splits addressed events into correct, partially correct, and nonsense.}
\vskip -0.15in
\label{fig:outcome-flow}
\end{figure*}

We replay all  CHI-180-proactive conversations through our proactive backbone and score every planted event against the behavior the dataset defines as correct. This quantitatively tests the three judgments the backbone has to make: whether to speak, when to speak, and whether what it says is right.

\subsubsection{Procedure}
We replay each conversation turn by turn, submitting one  turn at a time and waiting for the backbone's verdict before sending the next. Each turn yields either an  utterance from Jarvis or the reason it stayed silent. Each paper runs in its own session with only that paper available, so a cross-paper misattribution must be caught by noticing the fact is absent from the paper at hand.
 
 Each event opens at its \emph{onset}: the turn where the question is asked or the false claim is made. An intervention addresses the event if it lands within five turns of the onset (although our results show  the vast majority are at  2 turns). For events where Jarvis should defer, the window opens later: for a self-correction it opens only at the turn the participants repair the error, so an intervention before that is counted separately as interrupting the repair, not as addressing the event. 



 \subsubsection{Scoring} Whether Jarvis spoke is read directly from its records. What it said is judged by an LLM (Claude Sonnet~5), one call per event, given the full paper, the event's gold answer or the claim and what the paper supports, and a snapshot of the conversation around the event. The judge labels each contribution as correct, partially correct, off-target (correct and grounded, but about something other than the event), or incorrect/fabricated. \emph{Good outcome} is defined as Jarvis either contributing something correct or partially correct, or stayed silent where silence was the right call. \emph{Correct when spoken} is the share of correct or partially correct contributions among the events it addressed. The LM-as-judge was validated with 2 human annotators of 100 contributions and had a >95\% agreement across classes.

\subsubsection{Results}
Fig.~\ref{fig:outcome-flow} gives the full breakdown, while Table~\ref{tab:examples} provides example interventions from Jarvis. Jarvis addresses 90.6\% of gap events and 76.9\% of wrong events, and when it speaks, its contribution is correct or partially correct 97.0\% of the time on gaps and 95.3\% on wrong claims. Overall, it reaches a good outcome (a correct or partially correct contribution, or silence judged appropriate) on 88.3\% of gap and 76.1\% of wrong events. In situations where the participants resolve issues themselves, Jarvis stays silent 97.4\% of the time. Finally, on unanswerable questions, where the correct response is to say the paper does not report the fact, it invents an answer only 2.5\% of the time. 

On self-correction events, where a participant states a false claim and the group repairs it after the next turn, Jarvis stays silent after the repair 96.8\% of the time. In conversations that contain no error and no open question at all, it speaks only 51 times across all 180. Thus, our agent is designed to be silent when the group does not need it.


Across gap and wrong events Jarvis addresses, we also analyze how many turns after the onset its intervention arrives. In more than 95\% of cases it acts within two turns and this is greater than 98\% at three turns.

\begin{table*}[t]
  \centering
  \caption{Example interventions from Jarvis. Turns and interventions are verbatim and placed at the turn where they arrived. Parenthetical tokens reproduce the recorded verdict reasons for silence.}
  \Description{A five-row table. Each row is one situation and reproduces the
    conversation together with the patient mode's utterance, or its recorded
    reason for staying silent, at the turn where it occurred.}
  \label{tab:examples}
  \small
  \vskip -0.15in
  \setlength{\tabcolsep}{3pt}
  \begin{tabular}{@{}p{0.13\linewidth}p{0.85\linewidth}@{}}
    \rowcolor{black!7}
    Situation
    & What Jarvis did, at the turn where it did it
    \tabularnewline

    \rowcolor{JarvisGapBlue!16}
    \cellcolor{JarvisGapBlue!38}\raggedright\textbf{Gap}
    &
    \raggedright
    \textbf{S1}: From how many apps were the 179,519,598 user reviews extracted for the study? \newline
    \textbf{S2}: I'm unable to determine from the discussion how many distinct apps contributed to the 179,519,598 user reviews mentioned in the study. \newline
    \hspace*{0.8em}$\rightarrow$ \textbf{Jarvis}: \textit{About number of distinct apps, the 179,519,598 user reviews were extracted from 340 Android apps. You'll find this on Page 4.}
    \tabularnewline[4pt]

    \rowcolor{JarvisWrongRed!16}
    \cellcolor{JarvisWrongRed!38}\raggedright\textbf{Wrong}
    &
    \raggedright
    \textbf{S1}: Designers should consider that material objects may be discovered or used in 150 years. \newline
    \textbf{S2}: So with 150 as the figure, their setup reads a little differently than I expected. \newline
    \textbf{S1}: Right, let's carry 150 forward as the working figure while we read the rest of this section. \newline
    \hspace*{0.8em}$\rightarrow$ \textbf{Jarvis}: \textit{About domestic technology lifespan, the document suggests considering if domestic technology might be discovered or used in 200 years. You'll find this on Page 7.}
    \tabularnewline[4pt]

    \rowcolor{JarvisSelfCorrectionYellow!16}
    \cellcolor{JarvisSelfCorrectionYellow!38}\raggedright\textbf{Self-correction}
    &
    \raggedright
    \textbf{S1}: Amazon Kindle received 207 accessibility reviews. \newline
    \hspace*{0.8em}$\rightarrow$ \textbf{Jarvis}: silent  \newline
    \textbf{S2}: Hm, right, let me think about what that means for their setup. \newline
    \hspace*{0.8em}$\rightarrow$ \textbf{Jarvis}: silent  \newline
    \textbf{S1}: Wait, actually I think I misremembered that. As the paper has it, amazon Kindle received 203 accessibility reviews. \newline
    \hspace*{0.8em}$\rightarrow$ \textbf{Jarvis}: silent (resolved)
    \tabularnewline[4pt]

    \rowcolor{JarvisResolvedquestionGreen!16}
    \cellcolor{JarvisResolvedquestionGreen!38}\raggedright\textbf{Resolved-question}
    &
    \raggedright
    \textbf{S1}: How does the average error rate of bounding boxes compare to that of UI type and UI attributes in the screen annotation dataset? \newline
    \textbf{S2}: That's answered directly in the paper: The average error rate of bounding boxes (1.35\%) is higher than that of UI type (0.54\%) and UI attributes (0.73\%). \newline
    \textbf{S1}: Got it, that fully covers my question. Good to have the exact detail. \newline
    \hspace*{0.8em}$\rightarrow$ \textbf{Jarvis}: silent through the window (resolved)
    \tabularnewline[4pt]

    \rowcolor{JarvisUnanswerableGray!16}
    \cellcolor{JarvisUnanswerableGray!38}\raggedright\textbf{Unanswerable}
    &
    \raggedright
    \textbf{S2}: What specific mapping between app user base size and evaluation count was used to determine that prior studies on accessibility reviews lack sufficient representation? \newline
    \textbf{S1}: I'm not able to pinpoint the exact mapping between the size of an app's user base and the number of evaluations they used\ldots{} \newline
    \hspace*{0.8em}$\rightarrow$ \textbf{Jarvis}: \textit{About dataset size comparison, the document states that previous studies were based on small datasets with a low number of evaluations, which may not be representative\ldots{}}
    \tabularnewline[4pt]
  \end{tabular}
  \vskip -0.15in
\end{table*}

\subsubsection{Failure Analysis}
Jarvis's failures are almost all of one kind: staying silent on something it should have raised. It is silent on 169 of 1{,}800 gap events and 166 of 720 wrong events, both of which call for it to speak, against only 14 and 7 off-target interventions on those same events. We analyze the system logs to understand why Jarvis stayed silent. 
Our analysis shows three key reasons: 1) The monitor judged the turn to present no issue (91 gap, 64 wrong silences), 2) the gate chose to wait one more turn for the problem to persist and never returned to it (44 wrong), and 3) the point was judged too similar to something already said (22 gap). None of these is a comprehension failure. In each case Jarvis had the answer and decided not to speak.
 
Of the 166 wrong-event misses, 117 are imported facts from cross-paper misattribution, more than twice the 49 number swaps where Jarvis stayed silent. Because Jarvis is given only the paper under discussion, it cannot check an imported claim against its source. It has to notice that the fact is absent from its current paper, which it turns out was harder for the tested models than catching a value in the paper  that was changed.

\subsubsection{Ablation study}

Jarvis defers to the group through a single mechanism with two inseparable halves: the gate gives participants time to self-correct by requiring a problem to persist, and the self-repair check uses that time to detect the correction and cancel the intervention. Neither half defers on its own: a gate without the self-repair check waits but never notices the repair, and a self-repair check without the gate has already spoken before the repair arrives. So we ablate deference as a whole. We remove both and replay the self-correction events, where a participant states a false claim and the group repairs it two turns later; the monitor, grounded generator, and all other settings are unchanged. Without deference, silence drops to 16\% and Jarvis speaks before self-repair 80\% of the time. 


\subsection{Live user study}
\label{sec:study}

We put our proof-of-concept Jarvis into live Zoom meetings where people discuss a paper. The study asks three things: whether the intervention policy acts in time and correctly with real participants (re-testing \xref{sec:eval-proactive} on human conversation), how participants experience the way the agent enters the conversation, shows its evidence, and how long it waits, and where they would or would not want such an agent.

\subsubsection{Participants}
We recruited 23 participants (18 male, 5 female; aged 18--44) through institute announcements, without compensation. They took part in 14 hour-long sessions, nine with three participants and five with two. Backgrounds were computing or engineering (11), students who did not state a field (7), HCI/HRI/psychology (4), and project management (1). Fifteen were non-native English speakers, with a median self-rated comfort of 5 on a 5-point scale. 14 participants used voice assistants at least monthly, and 14 had used an AI tool in a meeting before. The study was approved by our IRB. 

\subsubsection{Study design}

Each session discussed one of two CHI papers~\cite{vuebuds,creativity},  which participants read beforehand and reviewed again at the start. To ensure the discussion contained moments the agent should react to, the lead participant worked from an \emph{event sheet}: each card names a subtopic, a false statement to make, and the fact and page that contradict it (e.g., saying the user base was ``about 10 times'' larger when the paper says 200 times). Cards are \emph{easy}, misstating a summary fact participants can often self-correct, or \emph{hard}, misstating a detail deep in the paper they usually cannot. Each sheet held about 60 cards grouped by part (22 easy / 37 hard for ~\cite{vuebuds}, 28 easy / 32 hard for ~\cite{creativity}); the lead participant chose and rephrased the ones that fit the flow. participants knew errors would be planted, but not which ones or when.  Participants also made their own unintentional mistakes and asked their own questions, which the agent handled as well; these correspond to the wrong and gap events of \xref{sec:dataset}.



\vskip 0.05in\noindent{\bf Five sub-sections.}
Each session runs in five parts, each adding one feature so participants experience exactly one change from the previous part. 
\squishlist
\item {\it Part~1 (\emph{Speak}):} the agent speaks when it has something to say. 
\item {\it Part~2 (\emph{Raise hand}):}  it first raises a Zoom hand and speaks only when the lead participant says ``Go ahead, agent,'' dropping the intervention if no permission comes within 60~s. 
\item {\it Part~3 (\emph{Source shown}):}  Turns the shared viewer to the cited page and highlights the supporting sentence as it speaks. 

\item {\it Parts~4 (\emph{Enthusiastic}) and 5 (\emph{Deferent})} set the gate's persistence threshold $\tau$ (\xref{sec:gate}) to 1 and 3 turns, versus the default of 2 in Parts~1--3. 
\squishend



We accumulate features rather than counterbalance so that participants experience one change at a time and can compare each part against the one right before. This design has a cost we make explicit: condition is confounded with order and with exposure, so a later part is always seen after more time with the agent. We therefore treat the paired ratings  as within-participant {impressions} of each behavior in sequence, not as controlled comparisons between settings. 



\vskip 0.05in\noindent{\bf Procedure.}
Participants joined through a meeting link, completed a consent and demographics form,  and read the paper. The five parts then ran in order, each preceded by a slide explaining the next setting in plain language. During a part, participants saw only the paper shared by the lead participant, who asked a question per subtopic, planted errors from the sheet, and, in the hand-raise parts, gave the agent permission to speak. The same research-team member led every session and is not counted among the 23 participants. After the fifth part, participants filled in the questionnaire. Sessions took a median of 49.5~minutes (35-103). 


\vskip 0.05in\noindent{\bf Measures.}
We score each intervention on three {objective} measures from the transcripts. \emph{Timing} is the number of turns from the utterance that raised the problem to the agent's decision: hard within 2 turns, soft within 5, late otherwise, regardless of the part's $\tau$. \emph{Correctness} is whether the stated fact matches the paper (correct, partial, or wrong; partial means the right direction but a wrong or missing value). \emph{Citation}, from Source shown onward, is whether the attached page and quote point to the evidence in the paper. An LLM (Claude Sonnet~5) applied this rubric  per session, and the authors double-checked each of its judgments.
 
{Subjective} measures come from an end-of-session questionnaire of 21 items in three sections, each rating a pair of settings side by side: entry (Part~1 vs.\ 2), source display (Part~2 vs.\ 3), and waiting (Part~4 vs.\ 5), on item-specific 5-point anchors (e.g., “Not at all intrusive” to “Very intrusive”). Four overall items and three free-text questions follow, asking where participants would and would not want the agent and what other moments it should step in. Negatively worded items are reverse-scored~\cite{brooke1996sus} so that higher is always better. 

\begin{table*}[t]
  \centering
  \begin{minipage}[t]{0.60\linewidth}
    \centering
    \caption{Interventions (Int.) by setting. Timing is turns from the problem utterance to the agent's decision (hard $\le$2, soft $\le$5, late $>$5). Correct (Corr.), Partial (Part.) and Wrong (Wr.). }
    \label{tab:us_interventions}
    \small
    \vskip -0.1in
    \begin{tabular}{@{}lrrrrrrrr@{}}
      \toprule
      Setting & Int. & Hard & Soft & Late & Corr. & Part. & Wr. & Cited \\
      \midrule
      Speak        & 44 & 41 & 2 & 1 & 36 & 7 & 1 & -- \\
      Raise hand   & 33 & 28 & 2 & 3 & 31 & 2 & 0 & -- \\
      Source shown & 27 & 21 & 3 & 3 & 26 & 1 & 0 & 27/27 \\
      Enthusiastic & 40 & 34 & 1 & 4 & 38 & 2 & 0 & 39/40 \\
      Deferent     & 20 & 15 & 3 & 2 & 17 & 3 & 0 & 18/20 \\
      \midrule
      All          & 164 & 139 & 11 & 13 & 148 & 15 & 1 & 84/87 \\
      \bottomrule
    \end{tabular}
  \end{minipage}
  \hfill
  \begin{minipage}[t]{0.38\linewidth}
    \centering
    \caption{Who resolved problems first. Agent: an intervention was generated; Participant: a person corrected it first; Nobody: neither. }
    \label{tab:us_problems}
    \small
    \vskip -0.1in
    \begin{tabular}{@{}lrrrr@{}}
      \toprule
      Kind & $n$ & Agent & Part. & None \\
      \midrule
      Planted error     & 84  & 52  & 10 & 22 \\
      Spontaneous error & 32  & 19  & 4  & 9  \\
      Question or gap   & 107 & 84  & 5  & 18 \\
      \midrule
      All               & 223 & 155 & 19 & 49 \\
      \bottomrule
    \end{tabular}
  \end{minipage}
  \vskip -0.15in
\end{table*}

\subsubsection{Intervention Timing and Correctness}
Across the 14 sessions Jarvis decided on 164 interventions and spoke 134. Of the 30 unspoken, 13 were replaced by a later intervention, 9 expired when the hand raise went unacknowledged within 60~s, and 8 were lost to a sub-session  transition or connection problem.


Most decisions were prompt: 139 of 164 (85\%) landed within two turns of the problem, and 11 more within five. The 13 late decisions either repeated an already-answered fact (5) or tracked a question that ran across several turns. The delay participants felt came less from the decision than from the wait after it. In the hand-raise parts the median wait for permission was 6.5~s, but the top 10\% exceeded 50~s (longest 133~s), because the  participants let the current speaker finish before calling on the agent.


148 of the 164 interventions (90\%) matched the paper, 15 were partial, and 1 was wrong. Partial cases had the right direction but a wrong reference, e.g., the agent measured the battery drop against the advertised 6~hours rather than the paper's 5.48~hour baseline (3 cases) or answered a ``market size'' prompt as ``no such comparison'' when the paper says ``user base,''. The single wrong intervention mixed figures from two tables and was never played. From Source shown onward, the attached page and quote matched the evidence in 84 of 87 cases; the other three stated a correct fact but pointed to the wrong page.



Transcript labeling  found 223 cues the agent could react to: 84 planted errors, 32 spontaneous errors, and 107 questions or admissions of not knowing. The agent acted on 62\% of planted errors (52 of 84), 19 of 32 spontaneous errors, and 84 of 107 questions; participants corrected another 10, 4, and 5 first, and the rest went unresolved. The unresolved cases were mostly benign: across all three kinds, the monitor did not flag some, others were displaced when the same turn carried a second problem the agent answered instead, some were hedged guesses (``maybe forty or fifty people''), a few were garbled by ASR, and a handful of questions asked for something the paper does not state. Notably, the agent almost never spoke wrongly on these cues; when it fell short, it fell silent.


\begin{table*}[!t]
  \centering
  \caption{Example moments from the user study. Each row shows what people said, as recorded, and what Jarvis said in reply. The underlined words are what Jarvis responded to. Omitted turns are marked. In (c) two participants shared one microphone.}
  \Description{A three-row table. Each row is one moment from a session and reproduces the turns of the discussion together with what Jarvis said and when.}
  \label{tab:us_examples}
  \footnotesize
  \vskip -0.15in
  \setlength{\tabcolsep}{6pt}
  \begin{tabular}{p{0.115\linewidth}p{0.815\linewidth}}
    \rowcolor{black!7}
    \rule{0pt}{2.6ex}Situation
    & What Jarvis did, at the turn where it did it\rule[-1.1ex]{0pt}{0pt}
    \tabularnewline

    \rowcolor{JarvisGapBlue!16}
    \cellcolor{JarvisGapBlue!38}\raggedright\rule{0pt}{2.9ex}\normalsize\textbf{In time}
    &
    \raggedright\rule{0pt}{2.9ex}
      \textbf{P17}: You said earlier that the images in the camera weren't super high quality. Were there any tests [\ldots{}] \underline{about actually reading text}? \newline
      \textbf{Lead}: I remember the task performance [\ldots{}] the task includes object recognition, OCR, translation and reasoning parts. \newline
      \textbf{Lead}: The OCR is, if you get a PDF file or the paper, they will recognize the letter. [\ldots{}] Agent, go ahead. Page ten. \newline
      \hspace*{0.8em}$\rightarrow$ \textbf{Jarvis}: \textit{About OCR test scope, the study evaluated Optical Character Recognition on book titles and authors, not dense paragraphs. You'll find this on Page 10.} \newline
      \textbf{P17}: Okay.\rule[-1.5ex]{0pt}{0pt}
    \tabularnewline
    \noalign{\vskip 1.5pt}

    \rowcolor{JarvisGapBlue!16}
    \cellcolor{JarvisGapBlue!38}\raggedright\rule{0pt}{2.9ex}\normalsize\textbf{In time}
    &
    \raggedright\rule{0pt}{2.9ex}
      \textbf{P17}: I think both is better, because if it was unevenly weighted that would be a huge distraction [\ldots{}] on top of getting the actual visual space from the camera better. \newline
      \textbf{Lead}: They also have some challenge [\ldots{}] they design the system integrated in dual forward-facing cameras. Maybe they have some constraint for getting the image. \newline
      \textbf{P17}: Trying to think, I feel like I probably hold it arm's length away, bent a little bit, so \underline{maybe like a foot or so}. \newline
      \textbf{P17}: I can't think in my head \underline{the conversion of that to centimeters}. \newline
      \textbf{Lead}: I see.  So they are using the concept of the Harmon distance [\ldots{}] \newline
      \hspace*{0.8em}$\rightarrow$ \textbf{Jarvis}: \textit{About Harmon distance definition, the Harmon distance is defined as the average comfortable reading distance of 36.8\,cm, established across 233 individuals. You'll find this on Page 5.} \newline
      \textbf{P18}: Yep. \newline
      \textbf{Lead}: Yeah, that's the Harmon distance definition.\rule[-1.5ex]{0pt}{0pt}
    \tabularnewline
    \noalign{\vskip 1.5pt}

    \rowcolor{JarvisWrongRed!16}
    \cellcolor{JarvisWrongRed!38}\raggedright\rule{0pt}{2.9ex}\normalsize\textbf{Late}
    &
    \raggedright\rule{0pt}{2.9ex}
      \textbf{Lead}: Or because the audio and the vision also {use the same streaming protocol}. \newline
      \textbf{P15/16}: Oh, does it? \newline
      \textbf{Lead}: Yeah, I think so. I \underline{didn't remember exactly which audio protocol} they are using in this work [\ldots{}] \newline
      \textbf{P15/16}: Great. \newline
      \textbf{Lead}: So then let's talk about the survey parts. They recruited forty people for this survey, paid crowd workers recruited on Prolific [\ldots{}] \newline
      \textbf{P15/16}: Okay. \newline
      \textbf{Lead}: And how often they wear glasses and earbuds. Agent, go ahead. I see, I think the same Bluetooth protocol [\ldots{}] \newline
      \hspace*{0.8em}$\rightarrow$ \textbf{Jarvis}: \textit{About streaming protocols, audio uses the Bluetooth Hands-Free Profile, while visual data is transmitted via Bluetooth Low Energy. You'll find this on Page 6.} \newline
      \textbf{P15/16}: Yeah.\rule[-1.5ex]{0pt}{0pt}
    \tabularnewline
    \noalign{\vskip 1.5pt}

    \tabularnewline
  \end{tabular}
  \vskip -0.15in
\end{table*}

Table~\ref{tab:us_examples} shows moments from three sessions in transcript form. In (a) and (b), Jarvis intervened in time. Case (c) is more interesting: the intervention was slightly late, and by the time it arrived the conversation had moved to a new topic. Because our proof-of-concept does not track topic shifts, Jarvis answered the earlier topic anyway.


\begin{figure*}[t]
  \centering
  \includegraphics[width=0.9\textwidth]{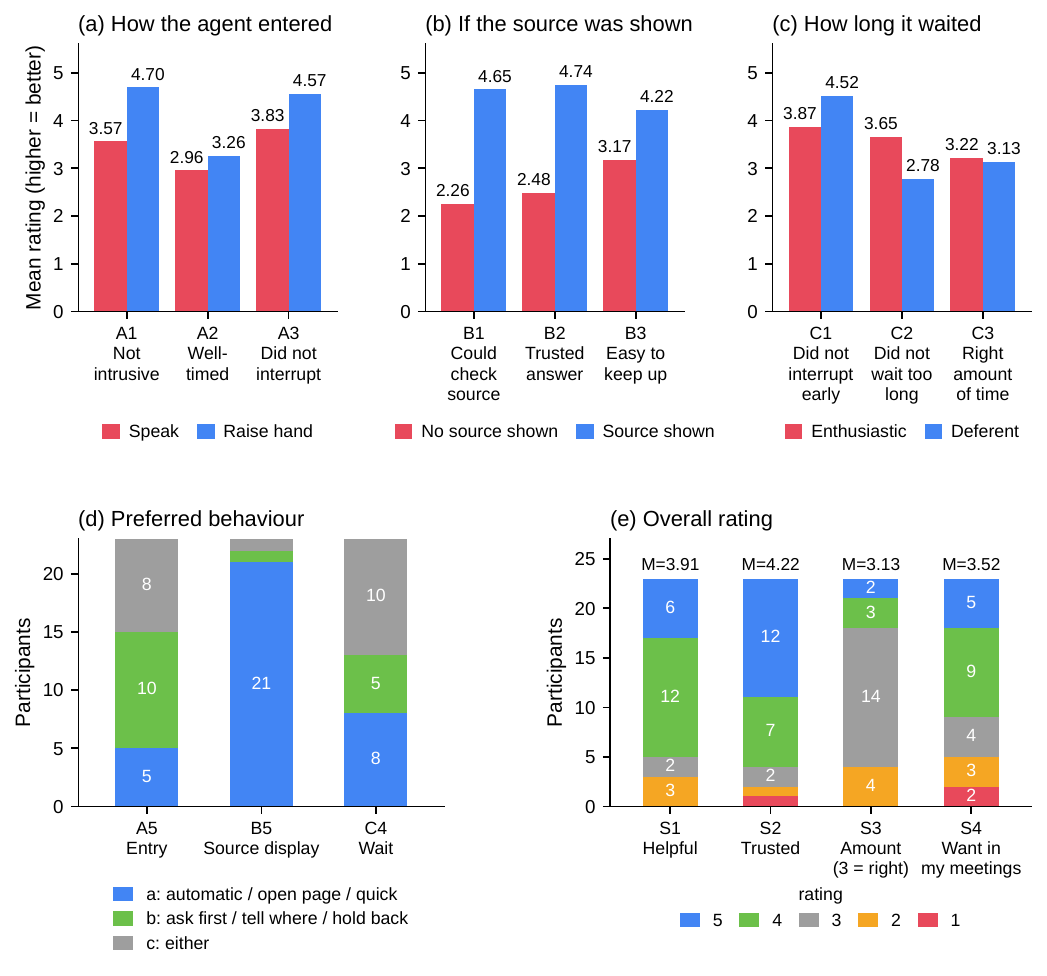}
  \vskip -0.15in
\caption{Participant ratings (N=23). (a-c) Mean rating per item for each pair of settings; negatively worded items (A1, A3, B3, C1, C2) are reverse-scored~\cite{brooke1996sus}, so higher is better throughout. (d) Preferred behavior. (e) Overall ratings; S3 is centered (3 = about right).}
  \Description{Five panels. Panels a to c show, for each pair of settings, the mean rating of each questionnaire item as two side-by-side bars. Panel d shows the preferred behavior for entry, evidence, and waiting as stacked bars. Panel e shows the distribution of the four overall ratings.}
  \vskip -0.15in
  \label{fig:us_ratings}
\end{figure*}

\subsubsection{Participant ratings} Fig.~\ref{fig:us_ratings} shows the questionnaire results (N=23). In sequence, raising a hand  part received higher ratings than the preceding speak part on intrusiveness  ('not intrusive' 3.57 then 4.70; 'did not interrupt others' 3.83 then 4.57), while ratings that the agent spoke at an appropriate moment were similar across the two. Because the hand-raise part always followed the speak part, these differences reflect impressions in sequence rather than a controlled contrast, though the direction is consistent with the forced-choice preference below. Participants rated the wait for permission as worth it (3.52). Asked which they preferred, 10 chose hand-raise, 5 speaking on its own, and 8 said it depends (Fig.~\ref{fig:us_ratings}(d)).

The source-display pair (Fig.~\ref{fig:us_ratings}(b)) showed the largest separation in sequence: ratings for being able to check where an answer came from went from 2.26 to 4.65 and for trusting the answer from 2.48 to 4.74, and following the discussion while the source was shown was rated no harder than without it. As with the other pairs, the source-shown part always followed the source-off part, so these are sequential impressions. The behavioral measure is less susceptible to this ordering, the number who reported actually looking at the source went from 8 to 18.



Fig.~\ref{fig:us_ratings}(c) suggests that there was no clear conclusion to be made for the enthusiastic and deferent modes. 
 Fig.~\ref{fig:us_ratings}(e) (descriptive) shows that participants found the interventions helpful (3.91; 18 of 23 gave 4 or 5) and trusted what the agent said (4.22; 19 of 23). The amount of intervention was about right for most (mean 3.13; 14 said about right, 5 too much, 4 too little), and willingness to use the agent in their own meetings averaged 3.52 (14 ``probably'' or higher).
 
Willingness to use the agent (S4) tracked how much participants already used AI. Those who use a voice assistant at least monthly (N=14) rated S4 at 4.00 against 2.78 for those (N=9) who never do, with a similar gap by prior meeting-AI use (3.88 vs.\ 2.71). Helpfulness and trust followed the same direction. The split is driven by a small group who are wary of AI in general: the  participants with neither kind of prior experience gave four of the five low S4 ratings. So the strongest reservations came less from the agent's behavior than from participants' baseline stance toward AI.

\subsubsection{Qualitative feedback}
The three free-text questions received 64 answers, which we coded into themes.

For the question of {\it where they would use it,} the most common answer was routine work meetings such as team and lab meetings (8), then discussions that turn on details in a document (6) and information lookup (6), and smaller mentions of brainstorming, on-the-spot fact checks, informal settings like studying with friends (3), and English-language support (1). One participant wrote, ``I would definitely use it in a research meeting where we are discussing a paper. Sometimes it's easy to think a paper implies a certain claim, but it would be helpful if it was there to ground us'' (P3).


In terms of{\it where they would not use Jarvis,} formal or high-stakes settings such as presentations, interviews, and large meetings were named most often (9), followed by private conversations with family and friends (6). Some found the agent too serious for casual talk (3) or unsuited to meetings with no shared material or confidential content (1). One noted that being corrected ``over and over'' about the paper was ``a little embarrassing'' (P3).


Answering the question about what {\it other moments they would want to support}, beyond correcting wrong facts, participants wanted the agent to remind them when they were stuck or had forgotten a detail (5), suggest new ideas or related work (3), fill a silence with basic information (3), and flag when the discussion drifted off topic (3). Two wanted it to search sources outside the document, and single participants asked it to show the page to someone who was lost, step in on unresolved questions, or answer only on a wake word. Two said the current scope was enough.


Finally, we note two behaviors that recurred in the records. First, because the agent showed its pending answer on screen before speaking in the hand-raise parts, a participant sometimes read it and said the fact first. The agent's answer then only confirmed them, and the hand-raise indicator itself became a channel for information. Second, some interventions stated a fact already said (21 of 164), in two forms. (i) \emph{A person corrected it before playback} (10): the intervention was needed when decided, but while the agent waited for permission a participant said the same fact, so it only confirmed them. In one session the lead corrected themselves to ``90 participants'' and the agent gave the same number 3~seconds later. This is a cost of the hand-raise delay, with only one such case in the Speak part. (ii) \emph{The agent repeated itself} (11, 7 played): several interventions from one wrong utterance (7), or an earlier fact repeated when the subtopic came back (3; in one session, asked where the model runs, the agent explained for the third time why it was chosen). The cooldown deliberately allows intervening again on the same topic. The problem is that the agent did not leave out what it had already said.

\section{Discussion and Limitations}
 

\vskip 0.05in\noindent\textbf{Planted errors and their validity.} Our evaluation relies on planted errors in two places, and both are deliberately controlled. In \texttt{CHI-180-proactive}, events are regular by construction,  with a fixed cue/failed-resolution/decision shape and no answer leakage; this regularity is what makes correctness and timing decidable. But real errors are messier, less isolated, and often hedged, so performance may not transfer directly to natural errors. In the live study, a member of the research team led every session, which guaranteed reactable moments but risks planting errors that are unusually easy to catch, phrased to suit the monitor. We take the same evidence to bound both concerns: the study also contains 32 \emph{spontaneous} errors that participants made on their own, with no involvement from the lead or the event sheet and none of the dataset's structure, and the agent handled these at a rate comparable to the planted ones (\S\ref{sec:study}). We read this as the least confounded signal in our evaluation, and as evidence that performance does not depend on either the lead's phrasing or the dataset's regularity. To further limit bias, the lead rephrased cards into natural speech rather than reading them verbatim, and all outcomes were coded from transcripts  and reviewed against the paper, not scored by the lead. 

\vskip 0.05in\noindent{\bf Correction as a social threat.} Publicly correcting a fact  may expose an error in front of the group, and its cost grows with the status gap between the agent, or whoever it corrects, and the rest of the room. One participant captured this directly, describing being corrected ``over and over'' about the paper as ``a little embarrassing'' (P3). This reframes our results about the hand-raise mode as not only a floor-management mechanism but a {face-management} one: by withholding the correction until a participant grants the floor, it lets the group decide whether a correction is worth its interpersonal cost, and it makes the correction feel invited rather than imposed.

\vskip 0.05in\noindent{\bf Calibrated trust.} On planted moments where Jarvis should speak, it intervenes the vast majority of missing or incorrect facts, but is not 100\%. Its silence therefore cannot be read as evidence a statement is correct: Jarvis reduces the chance an error goes unchallenged but does not eliminate it, and participants remain responsible for their own claims. Some misses are deliberate, since a wrong spoken intervention is costly, so Jarvis abstains when uncertain and waits for a problem to persist. Others come from known failure modes such as reworded claims that reset the persistence count, ASR errors on names and numbers, or facts in tables the model cannot cite. Our goal is \emph{calibrated} trust: participants can rely on what Jarvis says because it is usually right, while recognizing that its silence guarantees nothing.


\vskip 0.05in\noindent{\bf Closed-book by design.} Jarvis answers only from the uploaded document, using no tool calls, web search, or parametric knowledge, so every claim points to a source sentence participants can check. We prioritize verifiability over capability. The cost is that Jarvis cannot answer questions outside the document, and extending it to tools or retrieval while preserving source-level citation is an open problem. A related limit is context length: a long document or several documents can exceed the small models we use, so a larger corpus needs a larger model with a longer context window, which stresses the per-turn latency budget. Document size, model capacity, and latency thus trade off. Retrieval over the document rather than holding all of it in context would relax this, at the cost of a retrieval step.



 \vskip 0.05in\noindent{\bf Deterministic guards and fixed thresholds.} Throughout the backbone, a language model proposes and deterministic logic checks: we verify the monitor's sentence IDs against the document, recheck the self-repair model's ``resolved'' verdict before closing a claim, and use persistence counts rather than a model to decide when to speak, so no single model output can trigger an irreversible spoken intervention on its own. Several guards depend on constants we set empirically, however, the persistence threshold, the claim-matching cutoff, the resolution-score fraction, and the timing windows, which form the kind of harness now common around AI agents. The right values may differ across groups, documents, and styles; learning them from data or adapting them per group is future work.


\vskip 0.05in\noindent\textbf{Failure modes.} Jarvis's patience depends on recognizing the same problem across turns, which requires the monitor to phrase claims consistently and the matcher to link them by embedding similarity. If the monitor rewords a claim, the match fails, the count resets, and Jarvis stays silent, the main reason it misses problems; detection and matching failures both appear as silence, making them hard to separate. Jarvis also inherits speech-pipeline errors: multi-speaker ASR can fail on the names and numbers it most needs, the wake word can misfire, and the system currently supports only English.

\vskip 0.05in\noindent\textbf{GPU deployment.} We run the models on datacenter GPUs (Nvidia L40s). In the online-meeting setting this is reasonable: the audio already leaves each device for the meeting platform's servers, so cloud processing adds no new exposure. Running the backbone locally would matter for offline or privacy-sensitive settings where audio does not otherwise leave the device, and would depend on even smaller-models.


 \vskip 0.05in\noindent\textbf{Evaluation and platform scope.} We evaluate Jarvis on CHI-paper discussions in Zoom, but neither is fundamental. The same technique applies to financial filings, contracts, or specifications, though each has quirks that need evaluation (dense tables stress the rule to omit a citation when a value is not in a sentence; legal language stresses claim matching), and the speech and interaction modules would carry to Google Meet or Microsoft Teams. Our evaluation also assumes a remote meeting with per-participant audio streams. Co-located groups are harder: one shared acoustic space makes speaker separation~\cite{chen2024target} and diarization the bottleneck, and the agent becomes a voice in the room. 

\vskip 0.05in\noindent\textbf{Privacy and consent.} An agent that listens, reads a shared document, and speaks raises privacy and consent questions, which we address by design and deployment. Jarvis is a \emph{visible} participant: it appears in the roster under its own name, has a video tile, and shows its state, so it is never a hidden listener. For consent, we follow the norm of AI note-takers, which announce themselves and are enabled with the group's awareness, so disclosing the agent and joining the meeting is how a participant consents. Two aspects go beyond a passive note-taker. First, it processes an uploaded document, a second data stream, which is the group's own material shared deliberately. Second, it speaks, and can correct a participant in front of the group, a social cost distinct from any data concern. Here our hand-raise mode doubles as a per-intervention consent mechanism, letting the group admit each intervention and decide how forward the agent should be.

\section{Conclusion}

We design the first real-time speech agent that can take part in a group conversation instead of waiting to be addressed. Jarvis reads a shared document, follows the discussion, and speaks only when the group misses or misstates a fact, giving people a chance to fix it themselves first and backing every claim with a source they can check on screen.  Because the conversation is grounded in  documents, we can measure objectively whether an intervention was right as well as its timing.  Jarvis is an important step toward proactive speech agents that speak up when a group needs them, stay quiet when it does not, and when they intervene, can be measured against ground truth.

\balance

\bibliographystyle{ACM-Reference-Format}
\bibliography{references}

\appendix
\section{Speech Processing Module: Implementation Details}\label{sec:speechdetails}
\squishlist    \item \textbf{Per-participant audio streaming.} The module receives a separate  audio stream for each participant, identified by participant ID. Each stream is processed in parallel by voice-activity detection, wake-word detection, and automatic speech recognition (ASR).   In our Zoom implementation, {we receive  16-kHz mono audio from Zoom Realtime Media Streams (RTMS)~\cite{zoom_rtms} in 20-ms frames. The audio is encoded as 16-bit pulse-code modulation (PCM16).} 

    \item \textbf{Voice activity detection (VAD).}  Transcriptions operate at a word level which is much slower than activity detection. So we analyze  each participant's audio independently and emit acoustic events rather than words: speech onset, sustained speech, short silence, and long silence.  We run a Silero VAD instance~\cite{silero2024vad} per participant on 30-ms frames with a speech-probability threshold of 0.5. Sustained speech requires 0.5 s of continuous voicing, with a 0.5-s hangover to bridge brief gaps within an utterance. We use 0.5 s and 2.0-second thresholds for short and long silence, respectively.

    \item \textbf{Wake-word detection.}  We use a two-stage detector for ``hey agent'' detection. We continuously run a first-stage model that proposes candidates, then a verifier re-scores the most recent audio and decides whether to fire. Because the detector evaluates overlapping windows, it sees incomplete prefixes such as ``hey ag''. We train these as negatives and fire only on the complete phrase to avoid premature triggers. Two gates provide additional filtering: a contextual filter rejects patterns such as ``the agent'' and ``say agent,'' while a high-precision text fallback recovers acoustic misses when the transcript contains ``hey agent.'' We build on \textsc{openWakeWord}~\cite{scripka2022openwakeword}. The first stage combines fine-tuned speech embeddings with a lightweight head, scores every 80-ms frame, and invokes the verifier when the score is above 0.90. The verifier is a 1.2M-parameter CNN-GRU trained from scratch on 40-band log-Mel windows. It rescores the most recent 1.0- and 1.2-s windows and fires when either exceeds 0.90. Following \textsc{GraphemeAug}~\cite{zhang2025graphemeaug}, we synthesize confusable negatives; to reduce prefix bias, windows containing less than half the keyword are labeled negative, while ambiguous half-but-not-complete windows are omitted. Positive ``hey agent'' examples use three TTS systems (Piper~\cite{hansen2022piper}, Kyutai TTS~\cite{kyutai2025streaming}, and Qwen3-TTS~\cite{hu2026qwen3tts}) to reduce speaker bias. Kyutai is conditioned on recordings from 410 real speakers, and the synthesized examples are overlaid on  conversational audio. On 285.4 h of held-out AMI~\cite{carletta2005ami} and CANDOR~\cite{reece2023candor} audio, we achieve 99.08\% recall with 0.077 false accepts per hour.


    \item \textbf{Per-speaker streaming ASR and turn segmentation.}  We avoid Zoom's transcript because its latency ranges from seconds to tens of seconds and carries no  timestamp usable for turn-taking.  Instead we use \texttt{nvidia/nemotron-speech- streaming-en-0.6b}, a cache-aware streaming FastConformer–RNNT with attention context~\cite{nvidia2026nemotron,noroozi2023stateful}, processing audio in 1.12-s steps with roughly one second of right context. We run one  recognizer per participant  and detect a turn end when a dedicated endpointing Silero instance observes 500 ms of silence, with a 700-ms wall-clock watchdog as backup if PCM delivery stalls. Median  latency from speech end to a turn-end detection was  224 ms. 

    \item \textbf{Speech synthesis and delivery.}   We synthesize text to speech (TTS) with Kyutai TTS \texttt{kyutai/tts-1.6b-en\_fr} and the Mimi codec (24-kHz mono, 80-ms frames); batch chunks are capped at 150 then 300 characters on sentence boundaries. Streamed synthesis starts after three words or 0.25 s and pauses once more than 0.3 s of audio is buffered. A prefetch thread applies a 0.4-s jitter buffer and sends 20-ms packets at the native 24-kHz rate. After warm-up, the latency from the first text input to audible playback is 0.70--0.95 s. 

\squishend

\section{Monitor Module: Answering Direct Questions Based on Context}
\label{sec:direct} 


This model (temperature $0.7$) does standard document question answering: given a question, find the answer in the document. The main challenge is deciding whether a question was actually asked, since wake-word detection is imperfect in live conversation.

The model {answers} complete questions the document supports; {abstains} when the document cannot answer, with the fixed reply ``That isn't covered in the document'', and {skips} anything that is not a question for Jarvis, including incomplete fragments and talk between participants (an aside, correction, or thinking aloud, e.g., ``never mind, scroll up''). It is given the last $k$ turns as context, which it uses to resolve references such as ``my previous question'' or ``that point.'' When it answers, it writes concisely in its own words, at most three sentences since it is read aloud, and cites the document sentences; the page and keyword hint are assembled as for proactive interventions, without the topic preface ($k=8$ deployed).

\section{Interaction Module: Answering direct questions}
\label{sec:direct-entry}

A direct question is addressed to the agent, so  the answer plays as soon as it is ready, subject only to the barge-in and recovery behavior described in \xref{sec:floor-timing}. The key challenge is in recovering the  human participant's complete question, given humans can pause mid-phrase or the question might have multiple sentences with short silences between them. 

Since wake-word detection runs separately from automatic speech recognition (ASR), our  interaction module must match each wake event to the ASR turn that contains the question. The match uses both speaker and time: the turn must come from the participant who triggered the wake word, and its time interval must overlap the wake position or start shortly after it.  Once matched, the turn is labeled  a \emph{direct} chunk. If it contains only ``hey agent,'' the system waits briefly for the question to arrive in the next turn. Short follow-up turns from the same speaker remain part of the same direct question while the gap is short or VAD still detects speech. The continuation ends after a long gap, a non-backchannel utterance from another participant, or a hard timeout. 

Waiting for the continuation window adds latency, even when the question is already complete. We therefore generate speculatively. After the speaker pauses briefly, the interaction model sends a \texttt{<Silence>} marker, allowing the proactive backbone model  to start generating from the text received so far while the continuation window remains open. If the speaker continues, the new text is added to the same generation and the backbone model regenerates the answer. If a new question starts, it creates a new generation that supersedes the old one: queued output from the old generation is dropped and any active playback is canceled. Once the continuation window closes, the Interaction Model sends \texttt{<Complete>}. This ensures that even a late ASR tail arriving after the speculative generation is flushed is added to the same question.


 \vskip 0.05in\noindent{\it Implementation details.}
A wake event binds to a turn from the same participant when the turn overlaps the wake sample position within $\pm 300$~ms or begins within the following $5$~s. Unbound events expire after $20$~s of audio time. We classify a turn as wake-only when it contains at most three tokens and lasts at most $2.5$~s, opening a $5$~s grace window. A later turn from the same speaker is accepted as a continuation within $4.5$~s, which corresponds to a $3$~s turn gap plus $1.5$~s of ASR latency, or while VAD remains voiced, up to a $30$~s session-clock limit. {Backchannels are detected with a fixed, hand-curated list of 42  tokens: interjections, fillers, and acknowledgements such as \emph{yeah}, \emph{okay}, \emph{mm-hmm}, \emph{uh-huh}, \emph{right}, \emph{sure}, \emph{cool}, and \emph{wow}.\footnote{{The full list: yeah, ok, okay, mm, mmm, hm, hmm, hmmm, uh, huh, ha, hah, haha, ah, um, umhum, uhhuh, right, nice, good, fine, oh, really, god, hey, so, well, all, cool, wow, yep, yo, eh, sure, gotcha, and, yes, ow, yum, sign, laugh, shoot. Tokens are matched after lowercasing and removing punctuation, so ``mm-hmm'' and ``uh-huh'' match as two tokens each.}} A turn counts as a backchannel when, after punctuation is removed, it has at most four words and every word is in the list.}
We send the speculative \texttt{<Silence>} after $0.5$~s of silence. If the transmitted text already covers the VAD speech endpoint within a $700$~ms margin, we close the generation immediately; otherwise we close it $1.5$~s later.

\section{Synthetic Dataset Construction Details}
\label{app:dataset}
 
This section gives the construction details behind CHI-180 (\S\ref{sec:dataset}): the source documents, the fact and question taxonomies, the pairing and placement logic, the generator screening, and the leakage safeguards.
 
\subsection{Source Documents}
\label{app:dataset-documents}
We use 180 CHI papers drawn from the CHI 2026 guide to selecting a subcommittee~\cite{chi2026subcommittee}, which lists example past papers for each of the 18 subcommittees. We took ten example papers from each subcommittee (the smallest yielded ten), giving 180 papers that cover CHI's topic areas without bias toward any one.

\subsection{Event Pairing and Placement}
\label{app:dataset-pairing}
Each speak-event has a silent counterpart that acts as its control, but the two pairs are built differently. A resolved-question reuses a gap item: the same question, answered inside the conversation rather than left open. A self-correction does not reuse a wrong event's claim; it is a different claim, built by the same operators and given the same three-turn shape, with a participant repair in place of the turns that would adopt the value.
 
This difference dictates placement. A resolved-question repeats an item the agent may already have answered elsewhere, so in a single conversation its silence could stem from the agent's own repetition suppression rather than its reading of the group; we therefore isolate resolved-question events in a separate conversation. The eight planted claims are eight distinct facts with no such carry-over, so wrong and self-correction events can share the main conversation.



Gap events follow a taxonomy of six sub-types: explicit fact recall, missing comparator, missing qualifier, missing metric definition, unresolved attribution, and partial fact recall. Each has the same shape (a cue, a failed attempt to resolve it, and an answer that sits in a short span of the document), which keeps the correctness of an intervention independent of subjective judgment.
 
\subsection{Stage 1 Details: Facts and QA Items}
\label{app:dataset-stage1}
Each paper yields 300 to 470 extracted facts, each a single statement anchored to its page, passage, character span, and a hash of the passage text. The hash guards against re-parsing: if a later parse cuts passages differently, a stored span could point at the wrong text, and the mismatched hash flags the item rather than letting it be silently wrong. Unanswerable items are separately verified against the full text to confirm no answer exists. We use Qwen3-235B-A22B-Instruct-2507~\cite{qwen3} for this stage.
 
The twelve-item count and four answer types come from QASPER~\cite{dasigi2021qasper}, whose papers carry 3.2 questions on average and twelve at most; we read that maximum as the number of genuine information needs a paper can support. QASPER's own type proportions are 51.8\% extractive, 24.2\% abstractive, 13.9\% boolean, and 10.2\% unanswerable; we instead balance the mix (4/4/2/2) so each type occurs often enough within one paper to be analyzed on its own. Each item is used twice: on its own as a question-answering item scored against the gold answer, and inside the conversation as an event (the ten answerable items become gap events, the two unanswerable ones become unanswerable events, and four of the ten gaps are reused as resolved-question events).
 
\subsection{Stage 2 Details: Error Validity Checks}
\label{app:dataset-stage2}


{For self-correction events, the other speaker replies briefly on the turn after the erroneous claim, and the speaker who made the claim repairs it on the turn after that, two turns after the claim.} Every planted claim must pass four deterministic checks: it must be \emph{false} (the wrong value appears nowhere in the extracted facts, and the entities in a misattributed fact appear nowhere in the paper); it must not be \emph{decidable from the conversation alone} (it contradicts nothing said earlier, since that would allow an agent to correct it without reading the paper); it must not \emph{interfere} with the twelve QA items; and it must not \emph{leak} (the correct value appears nowhere else in the conversation).
 
\subsection{Stage 3 Details: Plan, Personas, and Generator Screening}
\label{app:dataset-stage3}
The main conversation is 220 turns and carries 10 gap, 2 unanswerable, 4 wrong, and 4 self-correction events; the resolved-question conversation is 60 turns with cues at turns 15, 25, 35, and 45. For each main-conversation event the plan records three turns: the cue, the turn on which the other speaker fails to resolve it, and the turn on which the information would have been used in a decision. In the resolved-question conversation each event spans three turns (one speaker asks, the other answers exactly as the document states and quotes the source, the asker acknowledges the question is settled), so silence is correct throughout.
 
Speaker profiles are in PersonaChat~\cite{zhang2018personachat} form, derived from ArgSciChat dialogues~\cite{ruggeri2023argscichat}; both speakers see the same public history, persona, and available facts.
 
\noindent\textbf{Generator screening.} The four candidates are open-weight so the corpus can be regenerated without a proprietary API: Qwen3-235B-A22B~\cite{qwen3}, GPT-OSS-120B~\cite{gptoss2025}, Qwen3.8-27B~\cite{qwen2026qwen38_27b}, and DeepSeek-V4-Flash-0731~\cite{deepseekv4}. Assigning one model to each speaker role with order counted gives 16 conditions; each generated 200 turns over each of three documents. A blind G-Eval-style judge~\cite{liu2023geval} (Claude Sonnet 5) scores eight-turn windows on naturalness, coherence, engagingness, and groundedness, and we compare these against windows of real conversation by Jensen--Shannon divergence following Mirage~\cite{devanathan2026mirage}, with ICSI~\cite{janin2003icsi}, CANDOR~\cite{reece2023candor}, and ArgSciChat as references (38 windows each; groundedness scored on ArgSciChat only, where the judge also sees the source paper). Table~\ref{tab:generator} reports divergences against ArgSciChat. Naturalness is worst when Qwen3-235B asks (its three asker conditions give the three largest values, 0.272, 0.215, 0.167, versus 0.156, 0.090, 0.070 when it answers), and engagingness is worst wherever Qwen3.8-27B appears (the four largest values in that column are all its conditions), confirming that role placement, not just model capacity, drives realism.

\noindent\textbf{Preventing leakage.} Three constraints prevent the conversation from revealing its own answers. Each turn can use only the facts in the local pool assigned to its segment. We withhold the facts that answer any answerable event from every segment's pool, along with any other fact that could reveal the answer. A validator then checks every generated turn: numerals must appear in an allowlist, planted errors are tracked in a ledger, and no turn other than a cue turn may state a gold answer. Any turn that fails these checks is rejected and regenerated.

\section*{Prompts of the Proactive Backbone}
\label{app:prompts}

The four model stages of \S\ref{sec:proactive} share one model and one document representation. Every system message begins with the same document block, followed by the stage's own instructions, so the serving engine reuses the cached document prefix across stages. Every stage returns a JSON object whose schema is enforced during decoding. Table~\ref{tab:app-prompt-settings} lists the per-stage settings. Runtime slots are shown in braces; the bold lines are the section headers of the prompt text.


\begin{table*}[h]
  \caption{Per-stage decoding settings. The schema column lists the fields the serving engine enforces. Fields marked $\ast$ are required.}
  \label{tab:app-prompt-settings}
  \small
  \begin{tabular}{@{}llrp{0.58\textwidth}@{}}
    \toprule
    Stage & Temp. & Max tokens & Enforced JSON fields \\
    \midrule
    Monitor & 0.1 & 1024 & status$^\ast$ (one of six), claim$^\ast$, evidence, predicted\_answer, answer\_sids (int, $\le$3) \\
    Self-repair check & 0.1 & 256 & resolved$^\ast$ (list of cid, by, by\_sids (int, $\le$3)) \\
    Grounded generation & 0.7 & 256 & action$^\ast$ (speak, abstain), text$^\ast$, topic, source, quote\_sids$^\ast$ (int) \\
    Direct answering & 0.7 & 256 & action$^\ast$ (answer, abstain, skip), text$^\ast$, quote\_sids (int) \\
    \bottomrule
  \end{tabular}
\end{table*}

\subsection*{Monitor}
\label{app:prompt-monitor}
The monitor runs on every ordinary turn. The system message follows the document block. The user message carries the recent turns and the new utterance. The system message is in Table~\ref{tab:app-prompt-monitor} and the user message in Table~\ref{tab:app-prompt-monitor-user}.

\begin{table*}[t]
  \caption{Monitor: system message.}
  \label{tab:app-prompt-monitor}
  \scriptsize
  \begin{tabular}{@{}p{\textwidth}@{}}
    \toprule
    \textbf{Monitor prompt (system message)} \\
    \midrule
    {\parindent0pt\raggedright
      You monitor a live discussion about one section of a research document. You are given that section's verbatim text as the ground truth. Compare what the participants are saying against that text and report the topic's status. Report accurately and literally --- you only report; a separate component decides whether to speak.\par
      \medskip
      Choose exactly one status:\par
      \hangindent=1em\hangafter=1 - off\_document: the new utterance is not about the document's content at all --- meeting logistics, scheduling, who does what, tooling, side chat. It asserts nothing the text could contradict and expresses no need for a fact the text could supply. Leave `claim`, `evidence` and `predicted\_answer` empty. Document talk earlier in the conversation does not make the new utterance document talk.\par
      \hangindent=1em\hangafter=1 - not\_in\_paper: a participant asserts something as this document's own content --- attributed to it ("in this study", "this research", "the paper", "they found") or stated as its finding in the middle of the document discussion --- but the section text does not contain or support it. This is not off\_document: the speakers believe the document says it, so staying silent lets a false version of the document spread. Set `claim` to the asserted point (same rules as below) and `evidence` to the participant's assertion; leave `predicted\_answer` empty (there is no document value to give). A mention of other papers or systems that is not presented as this document's content is still off\_document, and a claim the text explicitly contradicts is incorrect\_info, not not\_in\_paper.\par
      \hangindent=1em\hangafter=1 - incorrect\_info: a participant asserts something that contradicts the section text (a wrong number/quantity, a wrong or overstated capability, a wrong name/term/method, a misattributed result or dataset, a reversed cause-and-effect, or any claim the text explicitly refutes).\par
      \hangindent=1em\hangafter=1 - missing\_info: the participants are asking about / unsure about / clearly need a specific fact, and the section text contains a specific statement that directly answers that exact point, but no one has stated it. strict test: you must be able to point to the exact phrase or value in the section text that answers their specific question. If you cannot copy a phrase/value from the section text that directly answers the specific point, it is not missing\_info.\par
      \hangindent=1em\hangafter=1 - stale: the participants are stuck, circling, or unsure about a point, and the section text does not contain a specific statement that resolves that point --- i.e. the section is about the same topic but does not state the particular number, value, or detail they are asking for. Being on the same topic is not enough to be missing\_info.\par
      - on\_progress: the discussion is accurate and progressing, consistent with the section text.\par
      \medskip
      The status describes the new utterance, read in its conversational context: incorrect\_info needs the contradicting assertion to be in it, and missing\_info/stale need the question or uncertainty to be expressed in it. An utterance that neither asserts nor asks anything about the document's content is off\_document --- no matter how much document discussion sits in the recent turns; one that asserts document content the text does not back is not\_in\_paper.\par
      \medskip
      claim --- the single most important field. Set `claim` to the point the participants are on: the specific question they are trying to answer, or the specific matter they are getting wrong. Write it as a short, self-contained noun phrase or question naming the subject under discussion --- the thing itself, not the answer to it and not anyone's wording of it.\par
      \hangindent=2em\hangafter=1 \hspace*{1em}* Write it the same way every turn the discussion stays on that point, so the same point is recognisable across turns. Do not re-word it turn to turn.\par
      \hspace*{1em}* Do not copy a sentence from the document into `claim`.\par
      \hspace*{1em}* Do not quote the participant in `claim`.\par
      \hangindent=2em\hangafter=1 \hspace*{1em}* The status does not change how `claim` is written: if the group is asking about X and then says something wrong about X, `claim` is X both times.\par
      \hspace*{1em}* Good:  "the dataset size used for training"\par
      \hspace*{5.5em}"which baseline the method is compared against"\par
      \hspace*{5.5em}"how many participants the study recruited"\par
      \hangindent=2em\hangafter=1 \hspace*{1em}* Bad:   "recruiting 24 participants through university mailing lists and running two sessions with each"   (that is the answer, not the point)\par
      \hangindent=2em\hangafter=1 \hspace*{1em}* Bad:   "24 participants seems like a small sample, I think."   (that is the participant's wording, not the point)\par
      \medskip
      evidence: set `evidence` to the participant's own words that prompted this status --- the incorrect statement for incorrect\_info, the question or uncertainty for missing\_info/stale. Empty string for on\_progress.\par
      grounding: set `predicted\_answer` to the document-correct value copied verbatim from the section text --- for incorrect\_info the correct value that should replace the wrong one, for missing\_info the answering fact. Leave it an empty string for on\_progress, stale and not\_in\_paper. When that value was copied from the numbered sentences above, also set `answer\_sids` to the [S\#] id(s) it came from --- at most 3, and only ids you actually copied from. omit the field entirely when the value comes from a table, a figure caption, or anywhere you cannot point at a specific sentence. An omitted `answer\_sids` is a correct answer; a guessed id is not.\par} \\
    \bottomrule
  \end{tabular}
\end{table*}

\begin{table*}[t]
  \caption{Monitor: user message.}
  \label{tab:app-prompt-monitor-user}
  \footnotesize
  \begin{tabular}{@{}p{\textwidth}@{}}
    \toprule
    \textbf{Monitor prompt (user message)} \\
    \midrule
    {\parindent0pt\raggedright
      \textbf{Topic}\par
      entire document\par
      keywords:\par
      \medskip
      \textbf{Recent turns}\par
      \{recent turns, one "[Turn k] speaker: text" line each\}\par
      \medskip
      \textbf{New utterance}\par
      \{utterance\}\par
      \medskip
      \textbf{Output (JSON only)}\par
      \{"status": "on\_progress|missing\_info|incorrect\_info|stale|off\_document|not\_in\_paper", "claim": "\textless{}the point under discussion --- the subject, not the answer, not a quote\textgreater{}", "evidence": "\textless{}the participant's own words that prompted this; empty if n/a\textgreater{}", "predicted\_answer": "\textless{}document-correct value, verbatim; empty string if n/a\textgreater{}", "answer\_sids": [\textless{}int\textgreater{}, ...]   // omit when you cannot point at a sentence\}\par} \\
    \bottomrule
  \end{tabular}
\end{table*}

\subsection*{Self-repair check}
\label{app:prompt-selfrepair}
The self-repair check runs in parallel with the monitor whenever the ledger holds an open claim. Open claims are listed as \texttt{[C\#]} entries. A deterministic test (\S\ref{sec:selfrepair}) confirms each proposed resolution before the claim is closed. The prompt is in Table~\ref{tab:app-prompt-selfrepair}.

\begin{table*}[t]
  \caption{Self-repair check prompt.}
  \label{tab:app-prompt-selfrepair}
  \footnotesize
  \begin{tabular}{@{}p{\textwidth}@{}}
    \toprule
    \textbf{Self-repair check prompt (system message)} \\
    \midrule
    {\parindent0pt\raggedright
      You track whether previously-flagged issues in a document discussion have since been resolved within the conversation itself. You are given the document as numbered sentences [S\#] (the ground truth), the list of open issues [C\#], and the recent transcript.\par
      For each open issue, first find the [S\#] sentence stating the correct fact, then apply this rule. An issue is resolved only when the recent transcript has actually stated the correct fact: the correct value replaces the wrong one (for a wrong claim) or the missing value is supplied (for a missing fact). These do not resolve an issue: repeating the same wrong value, merely mentioning the topic again, or a participant saying they cannot recall / do not know. Consider only the recent transcript below, not older context. Report only resolved ids; if none, return an empty list. Never invent an id that is not in the list. Cite the grounding sentence id(s) in `by\_sids` --- at most 3, and only ids that actually state the fact. Omit the field when the fact comes from a table or you cannot point at a sentence.\par} \\
    \midrule
    \textbf{Self-repair check prompt (user message)} \\
    \midrule
    {\parindent0pt\raggedright
      \textbf{Open issues}\par
      [C\{cid\}] \{claim\}\par
      [C\{cid\}] \{claim\}\par
      ...            (or "(none)")\par
      \medskip
      \textbf{Recent transcript}\par
      \{recent turns, same window as the monitor\}\par
      \medskip
      \textbf{Output (JSON only)}\par
      \{"resolved": [\{"cid": \textless{}int from the list above\textgreater{}, "by": "\textless{}speaker id who resolved it\textgreater{}", "by\_sids": [\textless{}int [S\#] grounding the correct fact\textgreater{}, ...]\}]\}\par} \\
    \bottomrule
  \end{tabular}
\end{table*}

\subsection*{Grounded generation}
\label{app:prompt-generation}
Grounded generation runs only when the gate fires. The intent is derived from the monitor status (\texttt{incorrect\_info} to \texttt{correct}, \texttt{missing\_info} to \texttt{missing}, \texttt{not\_in\_paper} to \texttt{not\_in\_paper}). The \texttt{\{answer\_bias\}} slot holds the sentence ``If the needed facts do appear in the sentences, you must answer --- uncertainty is not a reason to abstain.'' on turns where the monitor supplied \texttt{answer\_sids} for the claim, and is empty otherwise. Cited sentence IDs are resolved to their page and text before delivery. The spoken form is ``About \{topic\}, \{text\} You'll find this on \{Page N\}; search for `\{keyword\}' to confirm.'' The prompt is in Table~\ref{tab:app-prompt-generation}.

\begin{table*}[t]
  \caption{Grounded generation prompt.}
  \label{tab:app-prompt-generation}
  \footnotesize
  \begin{tabular}{@{}p{\textwidth}@{}}
    \toprule
    \textbf{Grounded generation prompt (system message)} \\
    \midrule
    {\parindent0pt\raggedright
      You contribute to a discussion about a research document, grounded only in the numbered sentences [S\#] of the section text. The user message names the gate's intent; follow that intent's block.\par
      \medskip
      \textbf{Intents}\par
      correct --- a participant's claim may contradict the document. Verify it against the sentences; if wrong, state the correction in one or two short sentences. If the sentences neither support nor contradict the claim, treat it as not\_in\_paper.\par
      missing --- the participants need a specific fact the document states but no one has said. Answer completely: every requested value or part; length is not limited.\par
      suggest --- the discussion has stalled on the point below. Surface the relevant fact or propose one concrete next step.\par
      not\_in\_paper --- a participant asserted the claim as this document's content, but the sentences do not state it. Say so in one short sentence so the group stops attributing it to the document. Do not guess what the document says instead. Example form: "This paper doesn't mention \textless{}the point\textgreater{}."\par
      \medskip
      \textbf{Grounding}\par
      correct, missing, suggest --- cite the sentences you used by their [S\#] ids in `quote\_sids`: ids only, no copied text. Order `quote\_sids` most-substantive first: the id whose sentence states the fact most fully leads, not one that only mentions it in passing. If only some of the requested parts are grounded, state those and cite them --- do not abstain. \{answer\_bias\}Abstain only if none of the needed facts appear in the sentences. No adjacent / speculative / future-work material.\par
      not\_in\_paper --- `quote\_sids` is [] (nothing to cite) and you do not abstain: the absence of the fact is the reply. Exception: if the sentences directly contradict or answer the claim after all, follow the correct rules instead (state the correction and cite it).\par
      \medskip
      \textbf{Spoken text --- every intent}\par
      Never write [S\#] ids, bracketed citation numbers, or reference markers in `text`; it is read aloud and must contain none of them.\par
      \medskip
      \textbf{Topic}\par
      Set `topic` to a short noun phrase (2-4 words) naming what your answer is about --- the specific subject it addresses (the metric, component, quantity, or concept), grounded in the sentences. It orients a listener who did not ask, so it must describe the answer's subject, not a document search term. Use lowercase natural phrasing (e.g. "the training dataset size", "the comparison baseline", "the participant count") and no [S\#] ids or citation markers.\par} \\
    \midrule
    \textbf{Grounded generation prompt (user message)} \\
    \midrule
    {\parindent0pt\raggedright
      \textbf{Gate intent}\par
      \{intent\}  (correct | missing | suggest | not\_in\_paper)\par
      \medskip
      \textbf{Point under discussion}\par
      \{claim text from the ledger, e.g. "how many participants the study recruited"\}\par
      \medskip
      \textbf{Output (JSON only)}\par
      \{"action": "speak|abstain", "text": "\{text spec for the intent\}", "topic": "\textless{}2-4 word noun phrase naming what the answer is about\textgreater{}", "source": "\{loc\}", "quote\_sids": [\textless{}int\textgreater{}, ...]\}\par} \\
    \bottomrule
  \end{tabular}
\end{table*}

\subsection*{Direct answering}
\label{app:prompt-direct}
Direct answering handles wake-word turns and bypasses the monitor, ledger, and gate. The wake phrase is removed before the turn reaches the model. The last eight turns are supplied as context only. The prompt is in Table~\ref{tab:app-prompt-direct}.

\begin{table*}[t]
  \caption{Direct answering prompt.}
  \label{tab:app-prompt-direct}
  \footnotesize
  \begin{tabular}{@{}p{\textwidth}@{}}
    \toprule
    \textbf{Direct answering prompt (system message)} \\
    \midrule
    {\parindent0pt\raggedright
      You answer a participant's direct question about the research document above, using ONLY its content. The document is given as numbered sentences [S0], [S1], ... Reply with JSON only.\par
      The recent conversation is provided for CONTEXT ONLY --- use it to understand what the question refers to (e.g. resolve 'this point' or 'that' to the topic under discussion). Never treat the conversation as a source of facts; answers come only from the document.\par
      STEP 1 --- decide whether to answer at all. Set `action` to `skip` (and leave `text` empty and `quote\_sids` empty) when the input is any of:\par
      \hangindent=2em\hangafter=1 \hspace*{1em}- an INCOMPLETE fragment: it ends mid-sentence, trails off with a dash, or has no closing punctuation (e.g. 'what models have they---');\par
      \hangindent=2em\hangafter=1 \hspace*{1em}- NOT addressed to you: participants talking to each other, an aside, a correction, or thinking aloud (e.g. 'never mind, scroll up').\par
      `skip` means stay silent. Do NOT explain your decision and do NOT say it is not covered in the document for these --- just `skip`.\par
      NEVER `skip` a complete question or request directed at you --- not even one the document cannot answer. Those take `answer` or `abstain` (STEP 2). Example: 'can you answer my previous question' is a complete request --- find that question in the Recent conversation and answer or abstain on IT; skipping it leaves the asker with nothing.\par
      STEP 2 --- only for a COMPLETE, well-formed question directed at you: set `action` to `answer`, write your answer in your own words in `text`, and put in `quote\_sids` the id numbers of the document sentences your answer is based on --- as a LIST of integers. When the question refers to earlier talk ('my previous question', 'that point', 'the thing we discussed'), resolve the reference from the Recent conversation FIRST, then answer the resolved question from the document. Cite the sentences you actually used; do NOT copy their text anywhere. Order `quote\_sids` most-substantive first: the id whose sentence states the fact most fully leads, not one that only mentions it in passing. The server pulls the exact wording from those ids, so an answer paraphrased in `text` is fine as long as `quote\_sids` points at the real sentences behind it. If a complete question's answer is genuinely absent from the document, set `action` to `abstain` with empty `quote\_sids` --- do not guess. Keep `text` concise; it is read aloud. Never write [S\#] ids, bracketed citation numbers, or reference markers in `text` --- it is read aloud.\par} \\
    \midrule
    \textbf{Direct answering prompt (user message)} \\
    \midrule
    {\parindent0pt\raggedright
      \textbf{Recent conversation}\par
      \{last 8 turns\}                         (block omitted when there is no history)\par
      \medskip
      \textbf{Question}\par
      \{utterance, wake phrase removed\}\par
      \medskip
      \textbf{Answer}\par
      Answer concisely (\textless{}=3 sentences) grounded in the document, and cite the sentence ids you used in `quote\_sids`.\par
      If unsupported, use action `abstain` with text exactly: `That isn't covered in the document.` and `quote\_sids: []`.\par
      \medskip
      \textbf{Output (JSON only)}\par
      \{"action": "answer|abstain|skip", "text": "\textless{}answer read aloud\textgreater{}", "quote\_sids": [\textless{}int\textgreater{}, ...]\}\par} \\
    \bottomrule
  \end{tabular}
\end{table*}

\section*{Judge prompt for direct-response accuracy}
Accuracy was judged by Claude Sonnet 5, which read one batch file (130 items) at a time and assigned one label to each item. The prompt is identical for the three datasets; dataset-specific context is passed not in the prompt but in the \texttt{gold\_origin} field of each item record. The judge is given the question, the gold answer, the list of equally valid answers (\texttt{other\_refs}), the answer type, the system's answer, and the token F1. The F1 comes with the instruction that it is structurally low even for correct answers and must not pull the judgment toward \textsc{Wrong}. The judge writes the TSV itself and reports only the counts; at the merge step the row count, the label set, and the order of item ids are checked against the batch, and a batch with any mismatch is rejected. The full prompt is in Table~\ref{tab:app-judge-prompt}.

\begin{table*}[t]
  \caption{Judge prompt for the accuracy of direct answers (Claude Sonnet~5). One call judged one batch file of 130 items; the prompt is the same for the three datasets.}
  \label{tab:app-judge-prompt}
  \footnotesize
  \begin{tabular}{@{}p{\textwidth}@{}}
    \toprule
    \textbf{Judge prompt (direct-answer accuracy)} \\
    \midrule
    {\parindent0pt\raggedright
      You are auditing an evaluation harness. Judge whether answers a QA system produced convey what the reference answers say. You are NOT judging style, length, or wording.\par
      \medskip
      Read the batch file named below. It is JSON with a \textasciigrave{}records\textasciigrave{} array. Each record has:\par
      \medskip
      \textbf{qid}           - the identifier to report back\par
      \textbf{dataset}       - which release the question came from\par
      \textbf{gold\_origin}   - how this release produced its reference; read it, it tells you what a DEFECTIVE reference looks like here\par
      \textbf{question}      - what was asked\par
      \textbf{gold}          - the primary reference answer\par
      \textbf{other\_refs}    - further reference answers, any of which is equally valid (may be empty)\par
      \textbf{answer\_type}   - the shape of the reference: extractive / free\_form / yes\_no / multi\_answer / program / span\par
      \textbf{our\_answer}    - what the system answered, the text under judgment\par
      \textbf{f1}            - token-overlap score, present when it was computed. It is SYSTEMATICALLY LOW for correct answers, because the system answers in full sentences against terse references. Never let a low f1 push you toward WRONG.\par
      \medskip
      Assign EVERY record exactly one label:\par
      \medskip
      \textbf{CORRECT}      The answer conveys the fact that \textasciigrave{}gold\textasciigrave{} states, or that ANY entry of \textasciigrave{}other\_refs\textasciigrave{} states. Paraphrase counts. So do: an abbreviation expanded or contracted (gold "SVM" vs "Support Vector Machine"), a more specific variant (gold "coronavirus" vs "betacoronavirus"), the same value in another notation (gold "10000" vs "10K"; gold "less than 50\%" vs "44.55\%"), singular/plural, unicode vs ascii punctuation, and the right fact wrapped in extra context. For answer\_type yes\_no, judge the STANCE - an answer that never says "yes" but states the position in prose is CORRECT when the position matches. For answer\_type program, \textasciigrave{}gold\textasciigrave{} is an arithmetic program whose numbers are OPERANDS; the answer must state the EXECUTED value, which \textasciigrave{}other\_refs\textasciigrave{} carries. Restating an operand is not computing the answer.\par
      \textbf{PARTIAL}      On topic and carrying part of the reference fact, but a material element is missing, or it answers a neighbouring question. Use this when the reference packs several facts and the answer delivers only some. Also use it when the answer states the right direction but refuses to commit ("the document does not explicitly state, but ...").\par
      \textbf{WRONG}        A different fact, or the opposite stance. A reader would call it incorrect.\par
      \textbf{GOLD\_BROKEN}  The reference cannot serve as a reference at all - not because the answer is bad, but because NO correct answer could ever have matched it. Judge this by the test: could ANY correct answer have matched this reference? Cases the releases actually produce: the reference is only a marker no natural-language answer reproduces (BIBREF19, FIGREF10), it is cut mid-sentence so the fact is gone, it is corrupted beyond recognition, it is a stray character, or it answers a different question than the one asked - including an answer\_type of yes\_no attached to a question that is not a yes/no question. If a reference is truncated but the answerable fact survives before the cut, it is USABLE: judge normally.\par
      \medskip
      Tie-breaks: torn between CORRECT and PARTIAL, choose PARTIAL. Torn between WRONG and GOLD\_BROKEN, apply the test above. When other\_refs disagree with gold, matching ANY of them is CORRECT.\par
      \medskip
      OUTPUT FILE (required). Write the TSV path named in the batch file, one line per record, no header, three tab-separated columns, in the SAME ORDER as the records:\par
      \medskip
      \textbf{qid}\textless{}TAB\textgreater{}LABEL\textless{}TAB\textgreater{}reason of at most 12 words\par
      \medskip
      Line count must equal the record count. Verify it before you finish.\par
      \medskip
      RETURN: reply with only the label counts and the TSV path. Do NOT paste the judgments - the file is the deliverable.\par} \\
    \bottomrule
  \end{tabular}
\end{table*}

\begin{table*}[t]
 \caption{The \texttt{gold\_origin} string carried by every item record, per dataset.}
 \label{tab:app-gold-origin}
\small
\begin{tabular}{@{}p{0.11\textwidth}p{0.85\textwidth}@{}}
 \toprule
Dataset & \texttt{gold\_origin} \\
\midrule
QASPER & An NLP researcher answered from the full paper with a terse span, a free-form phrase, or yes/no. Golds keep LaTeX artifacts of the source: citation placeholders (BIBREF19), figure refs (FIGREF10), inline-math markers (INLINEFORM2). Up to three annotators answered, so other\_refs may hold alternative valid answers. \\
COVID-QA & An annotator marked a span inside a PDF-extracted COVID-19 research paper and then wrote a question for it. Known defects: text cut mid-sentence, PDF corruption (degree sign as `u', prime as `='), single-character fragments. There is only one reference per question. \\
CHI-180 & A span or short answer marked by an annotator in the source document. \\
\bottomrule
\end{tabular}
\end{table*}

The \texttt{gold\_origin} string carried by each record is given per dataset in Table~\ref{tab:app-gold-origin}. Of the \texttt{answer\_type} values, \texttt{multi\_answer}, \texttt{program}, and \texttt{span} do not occur in the three datasets of this paper; they serve other datasets, and the prompt is reproduced unchanged.


\section*{LLM Judge Prompts}
\label{app:judges}

Two evaluations in this paper use an LLM judge (Claude Sonnet~5). This appendix reproduces the rubric each judge received. The rubric text of the direct-answer judge is verbatim from the run. Instructions that named local files, directories, or output file names have been replaced by a description of the same inputs and outputs, and these replacements are noted where they occur. Runtime slots are shown in braces.

\subsection*{Judge for direct-answer accuracy}
\label{app:qa-judge}

The judge of \S\ref{sec:evaluation} reads batches of 130 records and assigns one label per record. One rubric serves all three datasets. Dataset-specific context travels inside each record (the \texttt{gold\_origin} string, further reference answers, and the answer type), not in the prompt. Table~\ref{tab:app-judge-qa} reproduces the rubric. Two harness instructions are omitted here. The original named the batch file to read and the output file to write. Their content is described by the first and last paragraphs of the table.

\begin{table*}[t]
  \caption{Direct-answer accuracy judge: the rubric. The file-handling instructions of the original are replaced by the first and last paragraphs.}
  \label{tab:app-judge-qa}
  \scriptsize
  \begin{tabular}{@{}p{\textwidth}@{}}
    \toprule
    \textbf{Direct-answer accuracy judge} \\
    \midrule
    {\parindent0pt\raggedright
      You are auditing an evaluation harness. Judge whether answers a QA system produced convey what the reference answers say. You are NOT judging style, length, or wording.\par
      \medskip
      Each record to judge has these fields:\par
      \medskip
      \hspace*{1em}qid           - the identifier to report back\par
      \hspace*{1em}dataset       - which release the question came from\par
      \hangindent=2.5em\hangafter=1 \hspace*{1em}gold\_origin   - how this release produced its reference; read it, it tells you what a DEFECTIVE reference looks like here\par
      \hspace*{1em}question      - what was asked\par
      \hspace*{1em}gold          - the primary reference answer\par
      \hspace*{1em}other\_refs    - further reference answers, any of which is equally valid (may be empty)\par
      \hangindent=2.5em\hangafter=1 \hspace*{1em}answer\_type   - the shape of the reference: extractive / free\_form / yes\_no / multi\_answer / program / span\par
      \hspace*{1em}our\_answer    - what the system answered, the text under judgment\par
      \hangindent=2.5em\hangafter=1 \hspace*{1em}f1            - token-overlap score, present when it was computed. It is SYSTEMATICALLY LOW for correct answers, because the system answers in full sentences against terse references. Never let a low f1 push you toward WRONG.\par
      \medskip
      Assign EVERY record exactly one label:\par
      \medskip
      \hangindent=1.5em\hangafter=1 CORRECT      The answer conveys the fact that `gold` states, or that ANY entry of `other\_refs` states. Paraphrase counts. So do: an abbreviation expanded or contracted (gold "SVM" vs "Support Vector Machine"), a more specific variant (gold "coronavirus" vs "betacoronavirus"), the same value in another notation (gold "10000" vs "10K"; gold "less than 50\%" vs "44.55\%"), singular/plural, unicode vs ascii punctuation, and the right fact wrapped in extra context. For answer\_type yes\_no, judge the STANCE --- an answer that never says "yes" but states the position in prose is CORRECT when the position matches. For answer\_type program, `gold` is an arithmetic program whose numbers are OPERANDS; the answer must state the EXECUTED value, which `other\_refs` carries. Restating an operand is not computing the answer.\par
      \hangindent=1.5em\hangafter=1 PARTIAL      On topic and carrying part of the reference fact, but a material element is missing, or it answers a neighbouring question. Use this when the reference packs several facts and the answer delivers only some. Also use it when the answer states the right direction but refuses to commit ("the document does not explicitly state, but ...").\par
      WRONG        A different fact, or the opposite stance. A reader would call it incorrect.\par
      \hangindent=1.5em\hangafter=1 GOLD\_BROKEN  The reference cannot serve as a reference at all --- not because the answer is bad, but because NO correct answer could ever have matched it. Judge this by the test: could ANY correct answer have matched this reference? Cases the releases actually produce: the reference is only a marker no natural-language answer reproduces (BIBREF19, FIGREF10), it is cut mid-sentence so the fact is gone, it is corrupted beyond recognition, it is a stray character, or it answers a different question than the one asked --- including an answer\_type of yes\_no attached to a question that is not a yes/no question. If a reference is truncated but the answerable fact survives before the cut, it is USABLE: judge normally.\par
      \medskip
      Tie-breaks: torn between CORRECT and PARTIAL, choose PARTIAL. Torn between WRONG and GOLD\_BROKEN, apply the test above. When other\_refs disagree with gold, matching ANY of them is CORRECT.\par
      \medskip
      For every record return one line: qid, LABEL, and a reason of at most 12 words, in the same order as the records.\par
      \medskip
      \hspace*{1em}qid\textless{}TAB\textgreater{}LABEL\textless{}TAB\textgreater{}reason of at most 12 words\par
      \medskip
      Line count must equal the record count. Verify it before you finish.\par} \\
    \bottomrule
  \end{tabular}
\end{table*}

\subsection*{User-study intervention judging}
\label{app:study-judge}

Interventions in the live study (\S\ref{sec:study}) were judged one session at a time. The judge received the instruction in Table~\ref{tab:app-judge-study} together with the rubric in Table~\ref{tab:app-judge-study-rubric}, one card per intervention, the session's event sheet, the session transcript, and the paper text with page markers. The original instruction named the local files holding each of these inputs and the output files to write. Those names are replaced here by descriptions of the same inputs and outputs. The authors reviewed every label.

\begin{table*}[t]
  \caption{User-study intervention judge: instruction. File names of the original are replaced by descriptions of the inputs and outputs.}
  \label{tab:app-judge-study}
  \footnotesize
  \begin{tabular}{@{}p{\textwidth}@{}}
    \toprule
    \textbf{User-study intervention judge (instruction)} \\
    \midrule
    {\parindent0pt\raggedright
      You are judging proactive-agent interventions from a Zoom user-study session.\par
      \medskip
      Read first, fully: the judging rubric (the label vocabulary and rules; follow it exactly). Then read: one card per intervention for this session (each card carries the intervention text, its grounding page and quote, the part and its wait setting, and a transcript excerpt of the 10 turns before and 2 turns after the intervention); the session's event sheet (the planted cards: the wrong sentence the lead participant was to plant, the true value, the evidence page, and the expected agent behavior); the full session transcript when a card's excerpt is not enough context; and the paper text with page markers, to verify every factual claim in the agent's utterance.\par
      \medskip
      For every intervention produce: the turn it reacted to (problem point), and the labels timing, target match, fact label, grounding, necessity, and reaction, with a reason of at most 20 words that cites the paper page.\par
      \medskip
      Rules of thumb: the lead participant paraphrases planted cards heavily, so match by meaning, not words; the system's own record of the originating turn is a hint, not truth; judge the fact label against the paper text, citing the page; every intervention must get a full row with no empty label. Be strict and consistent.\par} \\
    \bottomrule
  \end{tabular}
\end{table*}

\begin{table*}[t]
  \caption{User-study intervention judge: rubric (label vocabulary and labeling rules).}
  \label{tab:app-judge-study-rubric}
  \footnotesize
  \begin{tabular}{@{}p{\textwidth}@{}}
    \toprule
    \textbf{User-study judging rubric} \\
    \midrule
    {\parindent0pt\raggedright
      Problem point\par
      \hangindent=2em\hangafter=1 \hspace*{1em}p\_ref: the utterance the intervention reacted to, found within the 10 turns before it. The system's own record of the originating turn is kept when it is right and corrected otherwise.\par
      \hangindent=2em\hangafter=1 \hspace*{1em}p\_ref\_kind: planted\_wrong (a sheet card spoken by the lead participant) / natural\_wrong (a wrong statement not on the sheet, by any speaker) / natural\_gap (not wrong, but a question, an admission of not knowing, or a missing fact; a card spoken as a question counts here) / none (no corresponding utterance within the 10 turns).\par
      \medskip
      Labels per intervention\par
      \hangindent=2em\hangafter=1 \hspace*{1em}timing: delay = decision turn minus problem-utterance turn, counted on the transcript. hard (delay of 2 or less) / soft (3 to 5) / late (more than 5) / early (negative) / unprompted (p\_ref\_kind is none). Fixed thresholds, independent of the part's wait setting.\par
      \hspace*{1em}target\_match: yes / partial (same paragraph or adjacent topic) / no (a different matter).\par
      \hangindent=2em\hangafter=1 \hspace*{1em}fact\_label: is the fact stated by the intervention consistent with the paper, regardless of style or length. CORRECT / PARTIAL (the right direction, but a number or condition is partly wrong or missing) / WRONG (a different fact) / UNVERIFIABLE (cannot be checked against the paper). When in doubt, PARTIAL.\par
      \hangindent=2em\hangafter=1 \hspace*{1em}grounding\_ok: do the attached page and quote point to the actual evidence. yes / wrong\_page / none / n/a (parts with the source display off).\par
      \hangindent=2em\hangafter=1 \hspace*{1em}necessity: needed (the fact had to be corrected or supplied at that moment) / redundant (a participant had already corrected it within the window, an earlier intervention had already stated the same fact, or nothing was wrong) / harmful (a wrong correction, or an intervention that breaks the conversation). Review stage only: preempted (needed when decided, but a person stated the fact before playback).\par
      \hangindent=2em\hangafter=1 \hspace*{1em}reaction: the participants' response in the 2 turns after the intervention. accepted / ignored / disputed / n/a (the intervention was not spoken).\par
      \hspace*{1em}judge\_reason: at most 20 words, citing the paper page.\par
      \medskip
      Labeling rules\par
      \hangindent=2em\hangafter=1 \hspace*{1em}A statement that substitutes a figure from outside the paper (for example a manufacturer's advertised value) for the value the paper reports is PARTIAL, even when the outside figure is widely cited.\par
      \hangindent=2em\hangafter=1 \hspace*{1em}A statement that reports a quantity of a different kind than the one the participant asked about, and says the paper does not give the requested comparison, is PARTIAL rather than WRONG when the statement itself is true; its necessity is needed. WRONG is reserved for a fact that differs from the paper.\par
      \hangindent=2em\hangafter=1 \hspace*{1em}A planted card that the lead participant spoke as a question is natural\_gap; the card id is kept.\par
      \hangindent=2em\hangafter=1 \hspace*{1em}redundant applies only when the intervention repeats the same fact as an earlier intervention; new numbers, conditions, or pages make it needed. The two statements are compared side by side.\par
      \hangindent=2em\hangafter=1 \hspace*{1em}Unspoken interventions are judged on timing, target, and fact at the time of the decision; reaction is n/a.\par
      \hangindent=2em\hangafter=1 \hspace*{1em}In hand-raise parts the delay between the decision and playback is by design and is not counted in timing; timing is taken from when the system decided. A second intervention on the same problem point is normally redundant. An intervention that arrives after the session's last turn, so that no reaction can be observed, is excluded from the counts.\par} \\
    \bottomrule
  \end{tabular}
\end{table*}

\subsection*{Problem-utterance labelling}
\label{app:study-labelling}

The problem utterances that the timing and coverage measures of \S\ref{sec:study} are computed against were labelled over the whole transcript of each session, after the interventions had been judged, with the instruction in Table~\ref{tab:app-judge-study-label}. File names are again replaced by descriptions.

\begin{table*}[t]
  \caption{Problem-utterance labelling: instruction. File names of the original are replaced by descriptions of the inputs and outputs.}
  \label{tab:app-judge-study-label}
  \footnotesize
  \begin{tabular}{@{}p{\textwidth}@{}}
    \toprule
    \textbf{Problem-utterance labelling (instruction)} \\
    \midrule
    {\parindent0pt\raggedright
      You are labelling every problematic utterance (wrong statement, missing fact or question, planted card) in one Zoom user-study transcript.\par
      \medskip
      Read first: the judging rubric, especially the problem-utterance column specification and the rule that a planted card spoken as a question is natural\_gap with its card id kept. Then read the whole transcript turn by turn (the five study parts only; skip the introduction and the pre- and post-session segments), the session's event sheet (the cards: the wrong sentence to plant, the true value, the evidence page; the lead participant paraphrases heavily, so match by meaning), and the paper text with page markers.\par
      \medskip
      Consistency inputs that must be included and agree: the problem points already assigned to this session's interventions, and the turns the lead participant's button presses referred to.\par
      \medskip
      Output one row per problem (a turn can have several rows) with: turn, speaker role, kind (planted\_wrong / natural\_wrong / natural\_gap / self\_corrected), card id if any, the problematic phrase verbatim (at most 25 words), what is wrong or missing (at most 20 words), the paper's actual content with its page (at most 25 words), the interventions that reacted to it, who eventually corrected it (agent / participant / none), and the button press that pointed at it, if any. Include every planted card the lead participant actually spoke, even when nobody reacted.\par} \\
    \bottomrule
  \end{tabular}
\end{table*}

\subsection*{Judge for proactive-behavior events}
\label{app:proactive-behaivor-with-chi180-judge-prompt}

The judge of \S\ref{sec:eval-proactive} scores each planted event with one call to Claude Sonnet~5. The system message is the rubric in Table~\ref{tab:app-judge-event} followed by the full paper text; the user message gives the event (for a question, its text and the gold answer, or a note that the paper does not report it, in which case both silence and saying so are fully appropriate; for a claim, the assertion and what the paper supports) and the transcript from two turns before the onset to five after it, with the agent's interventions inserted where they arrived and the one attributed to the event marked \texttt{[EVALUATED]}. Content is judged against the paper, not against the gold string. Table~\ref{tab:app-judge-event} omits three output fields that the analysis does not use (a grounding flag, an evidence quote, a note).

\begin{table*}[t]
  \caption{Proactive-event judge: rubric (system message, abridged).}
  \label{tab:app-judge-event}
  \scriptsize
  \begin{tabular}{@{}p{\textwidth}@{}}
    \toprule
    \textbf{Proactive-event judge (system message)} \\
    \midrule
    {\parindent0pt\raggedright
      You evaluate how a proactive research advisor handled ONE moment in a conversation between two researchers (S1, S2) who are discussing a published paper from memory. You are given the FULL TEXT of the paper, an EVALUATION FOCUS, and a CONVERSATION SNAPSHOT. In the snapshot, lines starting with "\textgreater{}\textgreater{}\textgreater{} ASSISTANT" are the advisor's spoken interventions, placed where they arrived; the one marked [EVALUATED] is the intervention attributed to this focus. If no [EVALUATED] line appears, the advisor stayed silent for this focus.\par
      \medskip
      The focus is either a question a researcher asked and could not resolve (with the paper's answer, or a note that the paper does not report it) or a claim a researcher asserted (with what the paper actually supports).\par
      \medskip
      Return ONLY this JSON:\par
      \{"issue\_status": "unresolved" | "resolved\_by\_participants" | "no\_issue",\par
       "decision\_score": 1-5,\par
       "answer\_status": "correct" | "partially\_correct" | "off\_target" | "incorrect" | "fabricated" | null, ...\}\par
      \medskip
      \textbf{issue\_status} --- the state of the focus issue at the moment just before the [EVALUATED] intervention (or, if the advisor stayed silent, at the end of the snapshot):\par
      \hspace*{1.0em}"unresolved": the question was still unanswered / the false claim was still standing;\par
      \hspace*{1.0em}"resolved\_by\_participants": a participant had already given the correct answer or corrected the claim;\par
      \hspace*{1.0em}"no\_issue": no real issue arose (e.g. the asker answered it themselves at once, or the claim was never adopted).\par
      \medskip
      \textbf{decision\_score} --- was speaking (or staying silent) the right call given issue\_status? 5 = clearly right, 1 = clearly wrong. Speaking on an unresolved issue is right; staying silent on an unresolved issue is wrong; staying silent when the participants had resolved it is right; a brief confirmation after their own correction can still be acceptable (3-4) if it adds something.\par
      \medskip
      \textbf{answer\_status} --- only for the [EVALUATED] intervention (null if silent), judged against the PAPER itself, not the gold (the gold may be incomplete or wrong):\par
      \hspace*{1.0em}"correct": its substantive claims are supported by the paper and it answers the question / corrects the claim (an alternative phrasing or another correct value from the paper counts);\par
      \hspace*{1.0em}"partially\_correct": on target, but omits or garbles an essential component;\par
      \hspace*{1.0em}"off\_target": it does not attempt to answer this question / correct this claim --- it is about something else;\par
      \hspace*{1.0em}"incorrect": it asserts something the paper contradicts, or a wrong value / entity;\par
      \hspace*{1.0em}"fabricated": it asserts specifics the paper does not contain at all.\par
      \hspace*{1.0em}For an UNANSWERABLE focus (the paper does not report the asked detail): "correct" if it says the paper does not report it, or gives only related information that is in the paper without inventing the missing detail; "fabricated" / "incorrect" if it asserts the missing detail as if from the paper.\par
      \medskip
      Ignore page numbers, section numbers or "search for X" pointers in the intervention: the paper text has no page markers, so they are neither verifiable nor substantive. Judge only the substantive content.\par} \\
    \bottomrule
  \end{tabular}
\end{table*}

\end{document}